\documentclass{aa}

\usepackage{amsmath, amsthm, amssymb, amsfonts}
\usepackage[T1]{fontenc}
\usepackage{graphicx}
\usepackage{natbib}
\usepackage{subfig}
\usepackage{cleveref}

\newcommand{\spaceI}{\hspace*{0.25cm}}
\newcommand{\spaceII}{\hspace*{0.5cm}}

\crefname{section}{§}{§§}
\Crefname{section}{§}{§§}
\crefformat{section}{§#2#1#3}
\bibpunct{(}{)}{;}{a}{}{,} 
\title{}
\author{}
\DeclareUnicodeCharacter{2212}{-}

\begin{document}
        
        \title{Gaia-DR2 extended kinematical maps}
        
        \subtitle{Part III: Rotation curves analysis, dark matter, and Modified
Newtonian dynamics tests}
        
        \author{\v{Z}. Chrob\'{a}kov\'{a}\inst{1,2}, M. L\'{o}pez-Corredoira\inst{1,2}, F. Sylos Labini\inst{3,4,5}, H.-F. Wang\inst{6,7,8}, R. Nagy\inst{9}}
        
        \institute{Instituto de Astrofísica de Canarias, E-38205 La Laguna, Tenerife, Spain
                \and
                Departamento de Astrofísica, Universidad de La Laguna, E-38206 La Laguna, Tenerife, Spain
                \and
                Centro Ricerche Enrico Fermi, Via Panisperna 89A, I-00184 Rome, Italy
                \and 
                Istituto dei Sistemi Complessi, Consiglio Nazionale delle Ricerche, I-00185, Roma, Italia
                \and
                Istituto Nazionale Fisica Nucleare, Unit\`a Roma 1, I-00185, Roma, Italia
                \and
                South−Western Institute for Astronomy Research, Yunnan University, Kunming, 650500, P. R. China
                \and
                Department of Astronomy, China West Normal University, Nanchong 637009, China
                \and
                LAMOST fellow
                \and
                Faculty of Mathematics, Physics, and Informatics, Comenius University, Mlynsk\'{a} dolina, 842 48 Bratislava, Slovakia
        }
        
        \date{Received xxxx; accepted xxxx}
        
        
        \abstract
        {Recent statistical deconvolution methods have produced extended kinematical maps in a range of heliocentric distances that are a factor of two to three larger than those analysed in the Gaia Collaboration based on the same data.}
        {In this paper, we use such maps to derive the rotation curve both in the Galactic plane and in off-plane regions and to analyse the density distribution.}
        {By assuming stationary equilibrium and axisymmetry, we used the Jeans equation to derive the rotation curve. Then we fit it with density models that include both dark matter and predictions of the MOND (Modified Newtonian dynamics) theory. Since the Milky Way exhibits deviations from axisymmetry and equilibrium, we also considered corrections to the Jeans equation.
        %
        To compute such corrections, we ran N-body experiments of mock disk 
        galaxies where the departure from equilibrium becomes larger as a function of the distance from the centre.}
        {The rotation curve in the outer disk of the Milky Way  that is constructed with the Jeans equation 
                exhibits very low dependence on $R$ and $z$ and it is well-fitted both by dark matter halo and MOND models.
                %
                The application of the Jeans equation for deriving the rotation curve, in the case of the systems that deviate from equilibrium and axisymmetry,
                introduces  systematic errors that grow as a function of the amplitude of the average radial velocity. 
                In the case of the Milky Way, we can observe that the amplitude of the radial velocity reaches $\sim 10\%$ that of the azimuthal one at $R\approx 20$ kpc. Based on this condition, using the rotation curve obtained from the Jeans equation to calculate the mass may overestimate its measurement.}
        {}
        
        \titlerunning{Gaia-DR2 extended kinematical maps III.}
        \authorrunning{\v{Z}. Chrob\'{a}kov\'{a} et al.}
        \maketitle

\section{Introduction}\label{ch1}
Substantial progress has been made in the study of the Milky Way rotation curve\ thanks to the application of a novel range of methods. Inside the solar circle, the tangent-point method has been applied by measuring spectral profiles of the HI and CO line emissions \citep{burton}. Another approach considers the radial velocity of an object, 
which requires that its distance be measured independently, for example, by trigonometric or spectroscopic determinations. For this purpose, there is a variety of objects can be adopted, such as OB stars and their associated molecular clouds \citep{blitz}, the thickness of the HI layer \citep{merrifield}, the red giant branch and red clump \citep{bovy_red_clump,huang_red}, classical Cepheids \citep{pont_cef,mroz}, and a number of others. Rotation velocities can also be determined by measuring proper motions: when these are provided by  Very Long Baseline Interferometry (VLBI) techniques, the rotation curve can be determined with high accuracy \citep{honma}. The combination of proper motions from USNO-B1 observations with the Two Micron All Sky Survey (2MASS) photometric data has also been used to determine the rotation curve \citep{martin_rot_curve}. A powerful tool for measuring the rotation curve of the Milky Way is the  VLBI Experiment for Radio Astrometry (VERA), which uses trigonometric determinations of three-dimensional positions and velocities of individual maser sources \citep{reid_vlbi,honma_vera}. 

An significant study was carried by \cite{bhattacharjee} to construct the rotation curve of the Milky Way from $\sim 0.2$ kpc to $\sim 200$ kpc by using a variety of disk and non-disk tracers. In analysing the velocity anisotropy parameter, they also estimated a lower limit for the Milky Way mass. Their work was continued by \cite{bajkova}, who combined circular velocities of masers at low distances with the rotation curve of \cite{bhattacharjee} and fit the result using a number of models, varying, in particular, the dark matter halo, where they refine parameters for six different models. A comparison of some of our fit parameters with the results of \cite{bajkova} is given in \cref{ch5}. An excellent review of the current status of the study of the rotation curve of the Milky Way is given in \cite{sofue_review}. \\ 

Today, the Gaia mission of the European Space Agency
\citep{gaia2} provides a new possibility for studying the Milky Way with unprecedented accuracy thanks to data that offers the most accurate information about our Galaxy to date. Indeed, the Gaia data offer very precise determinations of position, proper motions, radial velocity measurements, and distance for millions of stars, although the errors of distance measurements increase with the distance from the observer. \\ 

In this paper, we present a systematical analysis of the Milky Way rotation curves derived by means of different methods and by using the Second Data Release (DR2) of the Gaia mission \citep{gaia}. To calculate the rotation curve, we use the Jeans equation that relates the circular velocity to observational quantities, such as the Galactocentric radial and tangential velocities, along with their respective dispersions. 
To do so, we must assume that the gravitational potential of Milky Way is axisymmetric and that the Galaxy is in a steady state configuration. 
In addition, by using numerical N-body experiments of simple disk models, we try to quantify the effect of the deviations from the equilibrium configuration on the determination of the rotation curve through the Jeans equation.

This paper is organized as follows: in \cref{ch2}, we describe the selection of the data used in this paper and in \cref{ch3}, we illustrate the method used to measure the Milky Way's rotation curve and present our determinations. In \cref{ch4}, we explain the method for calculating the density distribution from the Poisson equation by using the measured rotation curve. In \cref{ch5}, we fit different density models to our determination of the rotation curve using standard dark matter approaches, that is, by assuming that the Galaxy is embedded in a quasi-spherical halo whose mass can be then derived on the basis of such an hypothesis. In §6, we present our density models based on the Modified Newtonian Dynamics (MOND) theory. We study in \cref{ch6} the deviations from the  Jeans equation in out-of-equilibrium systems. Finally, in §8, we present our conclusions.

\section{Data selection}\label{ch2}
\citet[hereafter LS19]{martin} have produced extended kinematic maps of the Milky Way by using data from the second Gaia data release DR2 \citep{gaia} and considering stars with measured radial heliocentric velocities and with parallax error less than $100\%$. Their total sample contains 7 103 123 sources. Such objects were observed by the Radial Velocity Spectrometer \citep[RVS,][]{cropper}, which collects medium-resolution spectra (spectral resolution $\frac{\lambda}{\Delta \lambda}\approx 11 700$) over the wavelength range of 845-872 nm, centred on the Calcium triplet region. Radial velocities are averaged over a 22-month observational time span. Most sources have a magnitude brighter than 13 in the $G$ filter. \\

As the parallax error grows with the distance from the observer, LS19 applied a statistical deconvolution of the parallax errors based on the  Lucy's inversion method \citep{lucy} to statistically estimate the distance. In this way, they derived the extended kinematical maps in the range of Galactocentric distances up to 20 kpc. We chose this method due to its advantage over other Bayesian methods \citep[e.g.][]{bayes1,bayes2} as it does not assume any priors about the Milky Way density distribution. Any other method, such as the Lutz-Kelker method \citep{lutz}, is not appropriate here since it would  assume a uniform stellar volume density and a constant ratio $\sigma_\pi/\pi$, where $\pi$ is the observed parallax and $\sigma_\pi$ its standard deviation. For more details on this topic, see \cite{luri_paralaxy}, which gives an extensive analysis of different methods if inferring distance from the parallax, along with their respective advantages and disadvantages.

In further detail, the effective temperatures for the sources with radial velocities that LS19 considered
are in the range of 3550 to 6900 K. The uncertainties of the radial velocities are: 0.3 km/s at $G_{RVS} < 8$, 0.6 km/s at $G_{RVS}=10$, and 1.8 km/s at $G_{RVS}= 11.75$; along with systematic radial velocity errors of $< 0.1$ km/s at $G_{RVS} < 9$ and 0.5 km/s at $G_{RVS}= 11.75$. The uncertainties of the parallax are: 0.02 –0.04 mas at $G<15$, 0.1 mas at $G=17$, 0.7 mas at $G=20$ and 2 mas at $G=21$. The uncertainties of the proper motion are: 0.07 mas $\mathrm{yr}^{-1}$ at $G<15$, 0.2 mas $\mathrm{yr}^{-1}$ at $G=17$, 1.2 mas $\mathrm{yr}^{-1}$ at $G=20$ and 3 mas $\mathrm{yr}^{-1}$ at $G=21$. For details on radial velocity data processing and the properties and validation of the resulting radial velocity catalogue, see \cite{sartoretti} and \cite{katz}. The set of standard stars that was used to define the zero-point of the RVS radial velocities is described in \cite{soubiran}. LS19 consider the zero-point bias in the parallaxes of Gaia DR2, as found by \cite{lindegren,arenou,stassun,zinn}; however, they find that the effect of the systematic error in the parallaxes is negligible, so the maps that we use from their study (LS19, Figs. 8-12) do not consider the zero-point correction. We describe the way we use these maps in Sect. \ref{ch3} to construct the rotation curves, however, in Fig. 2, we include the zero-point correction to demonstrate that the difference is negligible.

\section{Rotation curves}\label{ch3}
From the Gaia DR2 catalogue, we estimate, for each object, the parallax $\pi$, the Galactic coordinates ($l,b$), the radial velocity $v_r$ , and two proper motions in equatorial coordinates $\mu_a\mathrm{cos}\delta$ and $\mu_\delta$. For our analysis, we need to know the Galactocentric position of stars in cylindrical coordinates ($R,z,\Phi$), and the Galactocentric velocity in cylindrical coordinates ($v_R,v_\Phi,v_z$). The transformation from these two coordinates systems can be found in LS19. \\

We limit the range of vertical distance to $\lvert z \rvert<2.2~$kpc as we find that far off-plane data are affected by larger errors in their parallax determinations.
We investigate the disk beyond the solar Galactocentric radius, that is, for $8.4~$kpc$<R< 21.2~$kpc. \\
To determine the rotation curve, we consider the
one component of Jeans equations in cylindrical coordinates \citep[Ch. 4.2, 4-29a]{binneyb}:

\begin{eqnarray}\label{1}
\frac{\partial(\nu\overline{v_{R}} )}{\partial t}+\nu\left(\frac{\overline{v_{R}^{2}}-\overline{v_{\Phi}^{2}}}{R}+\frac{\partial\Phi}{\partial R}\right)+\frac{\partial(\nu\overline{v_{R}^{2}} )}{\partial R}+\frac{\partial(\nu \overline{v_{R}v_{Z}})}{\partial z} = 0~,
\end{eqnarray}
{where $R$ is the Galactocentric radius, $v_R$ is the radial velocity, $v_Z$ is the vertical velocity, $v_\Phi$ is the azimuthal velocity, and $\nu$ is the volume density. The quantity $\overline{v^2}$ is the average square velocity for each component that can be written as $\overline{v^2}=\sigma^2+\overline{v}^2$, where $\sigma$ is the velocity dispersion. For a detailed calculation of the velocities and their respective dispersion, see LS19.}

\begin{figure*}
        \centering
        \subfloat[Rotation curves at different heights for positive values of $z$.]{
                \includegraphics[width=0.5\textwidth]{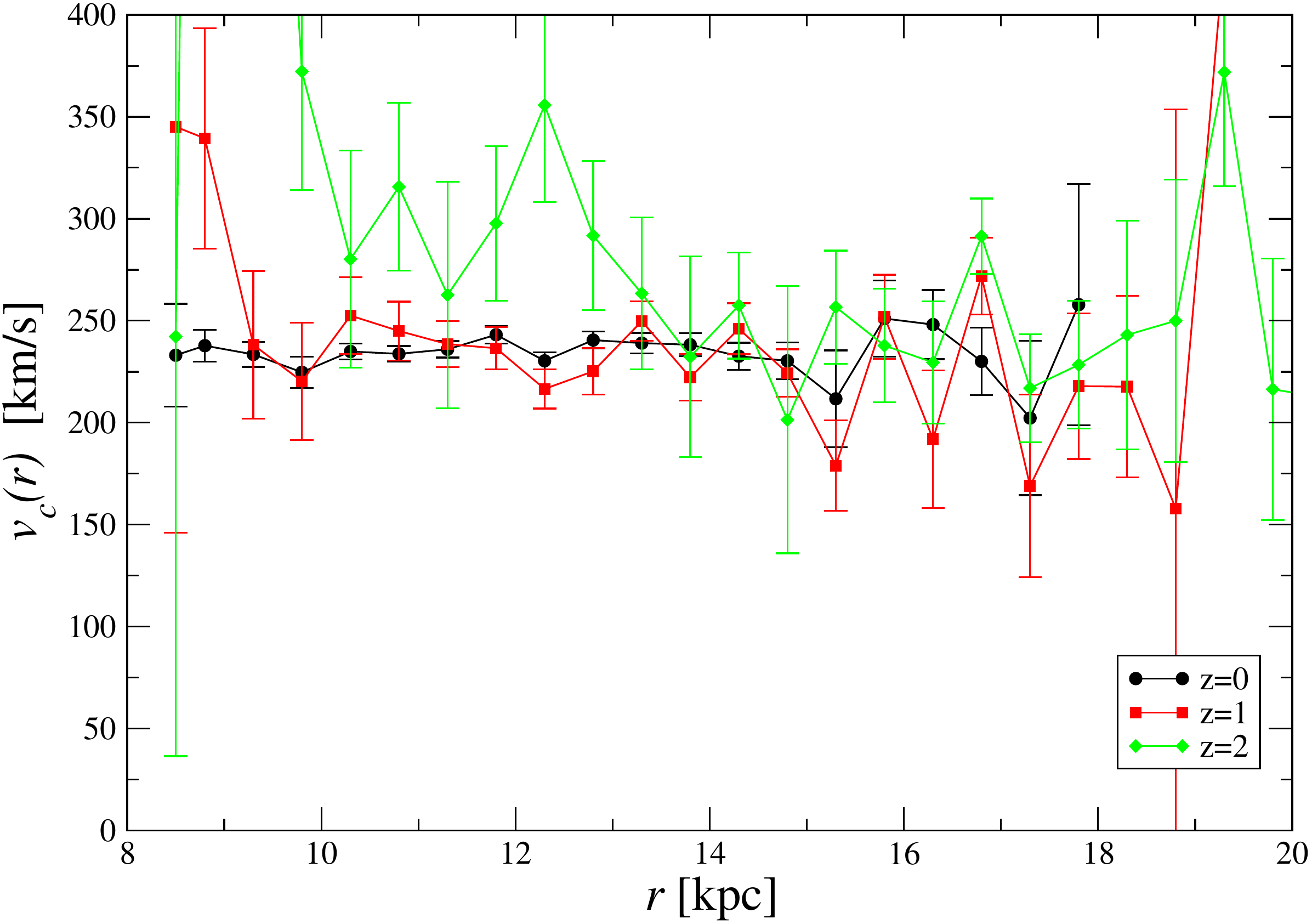}
        }
        \subfloat[Rotation curves at different heights for negative values of $z$.]{
                \includegraphics[width=0.5\textwidth]{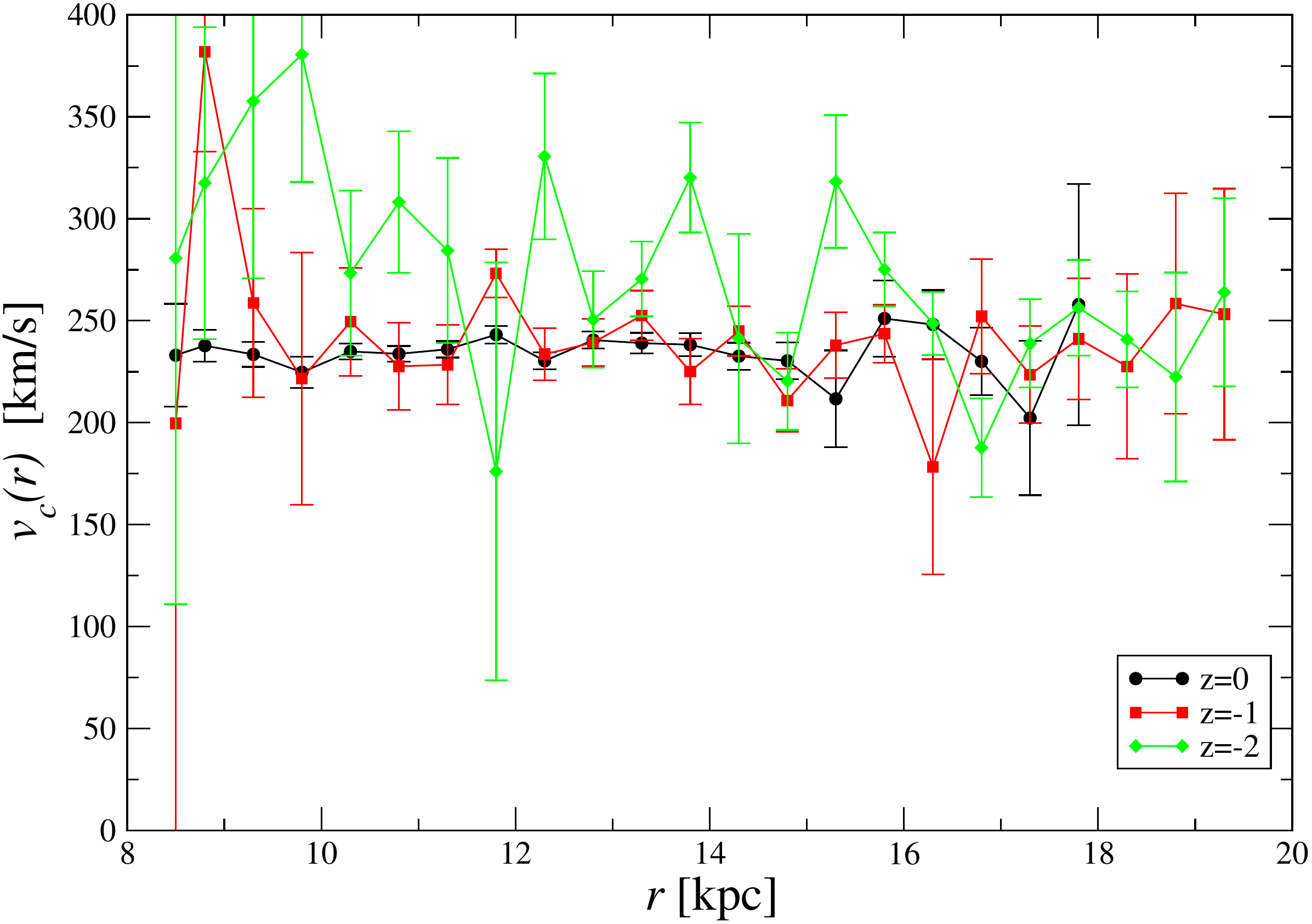}
        }
        \caption{Rotation curves for 
                different values of $z$. The error bars represent the standard deviation.} 
        \label{o1}
\end{figure*}

The rotational velocity is defined as \citep{binneyb}
\begin{eqnarray}\label{5a}
v_{c}^{2}(R,z)=R\frac{\partial\Phi}{\partial R}~.
\end{eqnarray}
We use the standard assumption that the volume density can be written as
\begin{eqnarray}\label{3}
\nu(R,z)=\rho_0e^{-\frac{R}{h_{R}}}e^{-\frac{\lvert z \rvert}{h_{z}}}~,
\end{eqnarray}
where $h_R$ is the scale length and $h_z$ is the scale height.
From Eqs. (\ref{1})-(\ref{3}) we obtain the rotational velocity as function of $R, z$, that is, the rotation curves,

\begin{eqnarray}\label{2}
v_{c}^{2}&=&\overline{v_{\Phi}}^{2}+\sigma_{\Phi}^{2}+\left(\overline{v_{R}}^{2}+\sigma_{R}^{2}\right)\frac{R-h_{R}}{h_{R}}-2R\overline{v_{R}}\frac{\partial\overline{v_{R}}}{\partial R} \nonumber \\
&-&R\frac{\partial \sigma_{R}^{2}}{\partial R}+\frac{R}{h_z}\frac{z}{\lvert z \rvert}\overline{v_{R}v_{z}}-R\frac{\partial (\overline{v_{R}v_{z}})}{\partial z}~.
\end{eqnarray}  
We determined the rotation curves for different values of $z$, in the direction of the anti-center, in bins of size $\Delta R=0.5$ kpc and $\Delta z=0.2$ kpc. 
For what concerns the scale parameters in Eq.\ref{3}, we chose values of $h_{R}=2.5$ kpc and $h_{z}=0.3$ kpc \citep{juric}.
Figure \ref{o1} shows the results of our fit. 

\begin{figure*}
        \centering
        \subfloat[Rotation curves at different heights for positive values of $z$.]{
                \includegraphics[width=0.5\textwidth]{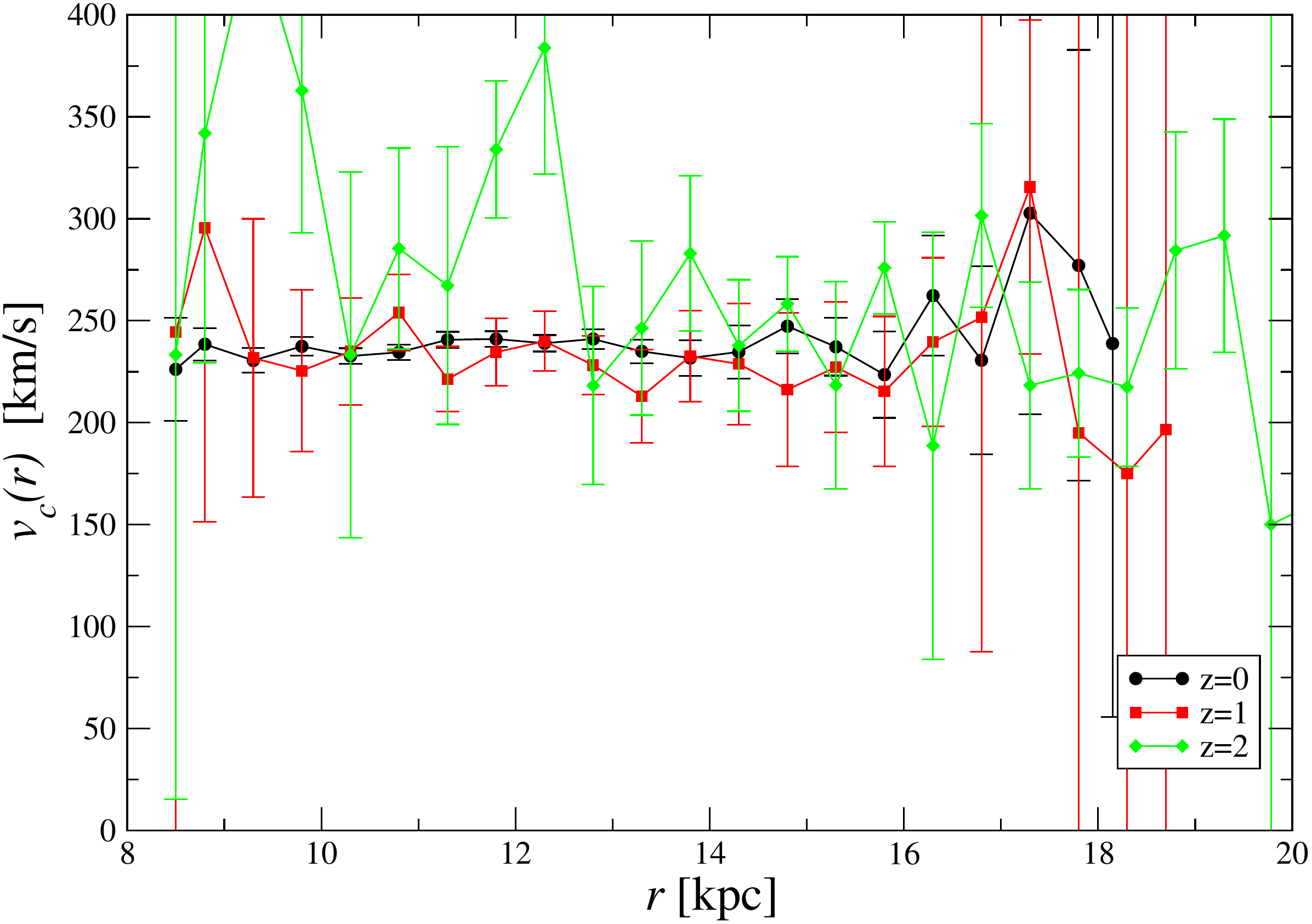}
        }
        \subfloat[Rotation curves at different heights for negative values of $z$.]{
                \includegraphics[width=0.5\textwidth]{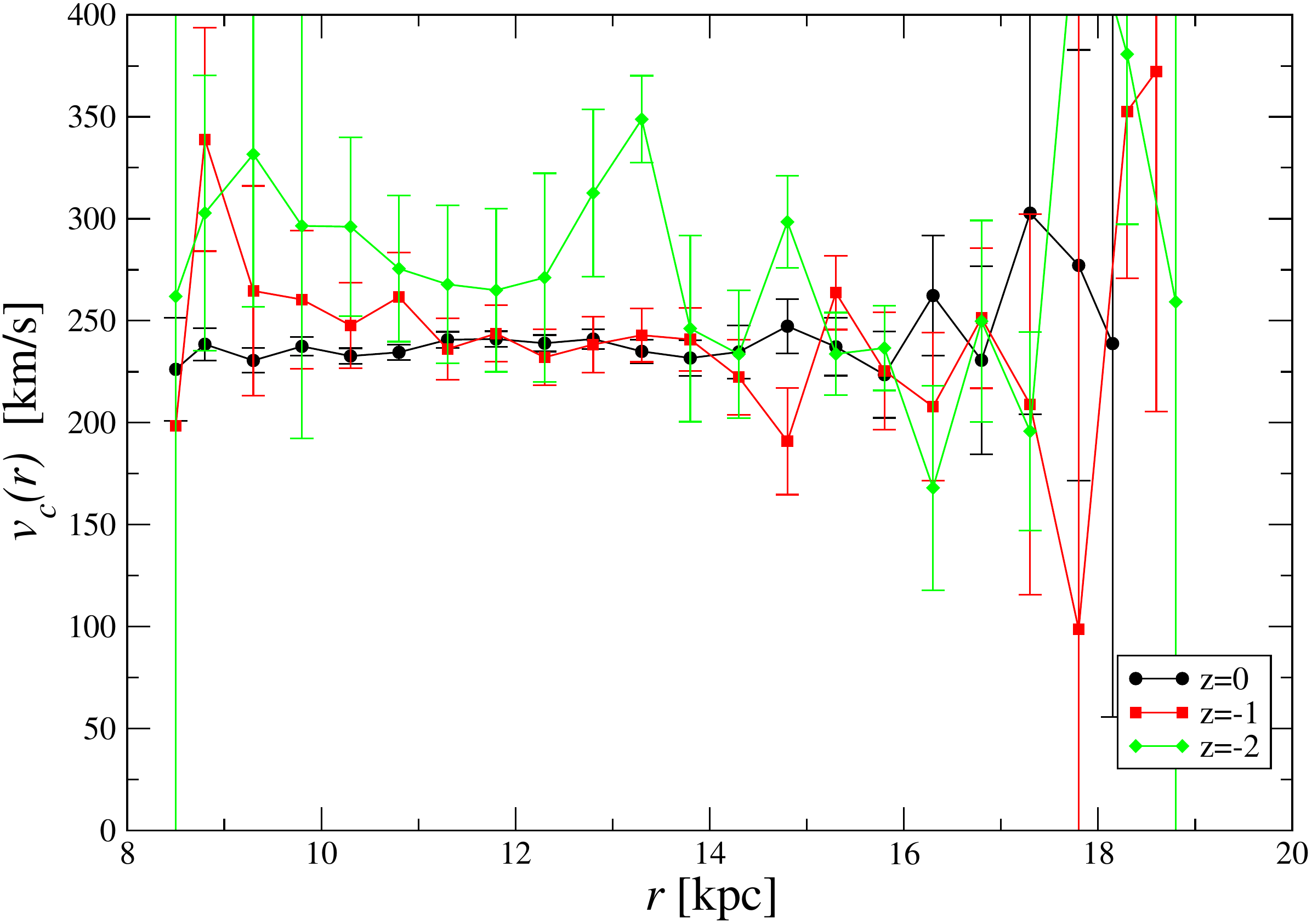}
        }
        \caption{Rotation curves for 
                different values of $z$ including the zero point correction in parallax. The error bars represent the standard deviation.} 
        \label{o1b}
\end{figure*}

\begin{figure*}
        \centering
        \subfloat[Rotation curves for different values of $h_{R}$, for $z=0$.]{
                \includegraphics[width=0.5\textwidth]{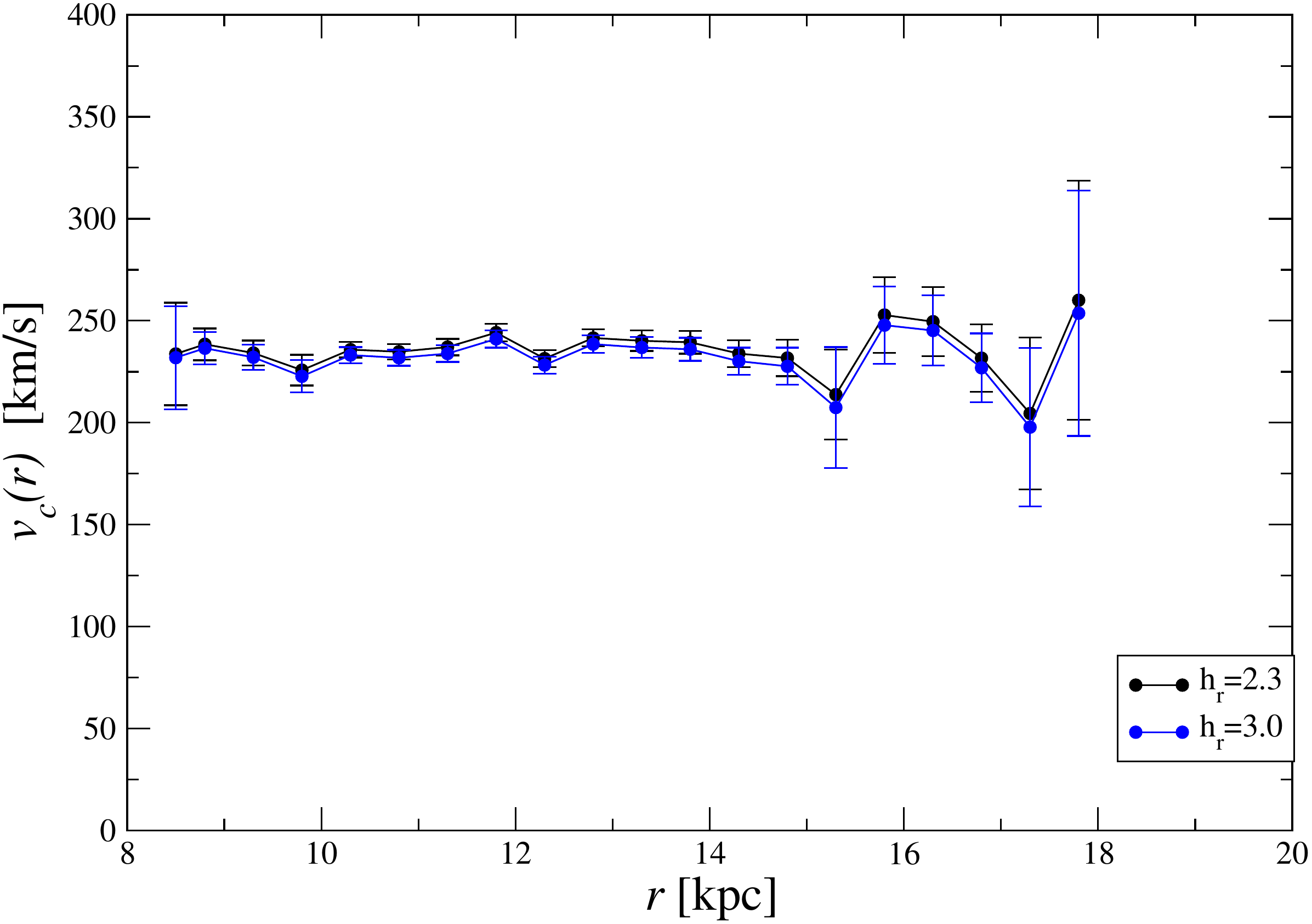}
        }
        \subfloat[Rotation curves for different values of $h_{z}$, for $z=0$.]{
                \includegraphics[width=0.5\textwidth]{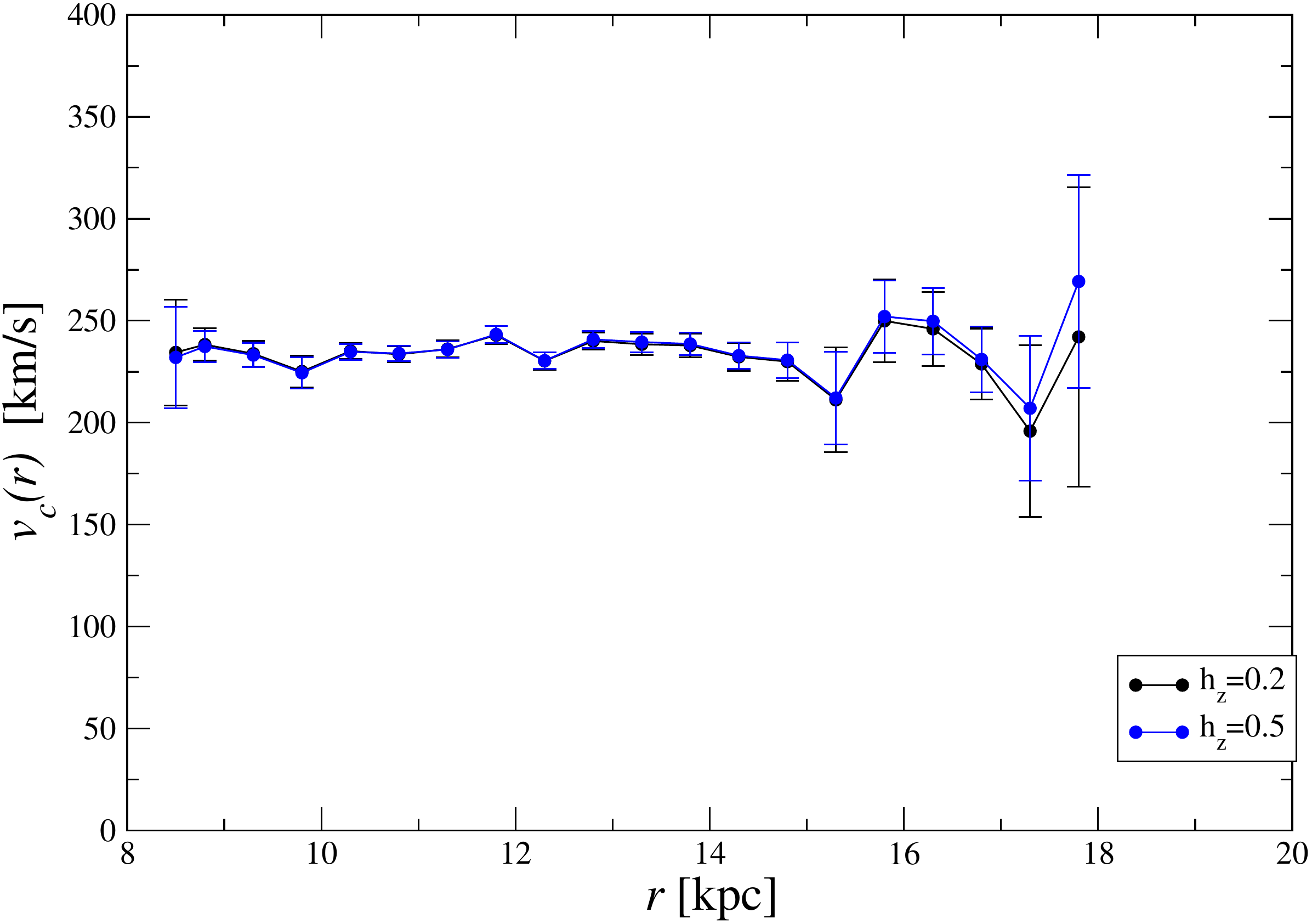}
        }
        \caption{Rotation curves for different values of the scale parameters.  The error bars represent the standard deviation.}\label{o2}
\end{figure*}

Figure \ref{o1b} shows the rotation curves, including the zero-point correction in parallax. The difference from the rotation curves in Fig. \ref{o1} is negligible and we do not consider this correction for the rest of the analysis. The results for different scale parameters are almost identical, as we show in Fig. \ref{o2}. We observe a flat rotation curve, although it does exhibit some fluctuations. Our rotation curve in the plane of the Galaxy has a small positive gradient of $0.54\pm0.7 (stat.) \pm 0.5 (syst.)~ \mathrm{km~s}^{-1}~\mathrm{kpc}^{-1}$. Recent results have shown an opposite trend: \cite{eilers} measured rotation curve for Galactocentric distances $5~$kpc$~\leq R \leq 25~$kpc by combining spectral data from the Apache Point Observatory Galactic Evolution Experiment \citep[APOGEE,][]{apogee} and photometric information from Wide-field Infrared Survey Explorer \citep[WISE,][]{wise}, 2MASS \citep{2mass}, and Gaia DR2, finding a rotation curve with a declining slope of $-1.7\pm0.1~\mathrm{km~s}^{-1}~\mathrm{kpc}^{-1}$, with a systematic uncertainty of $0.46 ~\mathrm{km~s}^{-1}~\mathrm{kpc}^{-1}$. A similar result was obtained by \cite{mroz}, who used classical Cepheids to obtain the rotation curve of the Milky Way for Galactocentric distances $4~$kpc$~\lesssim R \lesssim 20~$kpc, finding a rotation curve with a small negative slope of $-1.34\pm0.21~\mathrm{km~s}^{-1}~\mathrm{kpc}^{-1}$. \cite{bhattacharjee} have also used the Jeans equation, but only for large distances (R>20 kpc), which we do not consider in our analysis. For the disk tracers, they use the tangent point method for small distances and for higher distances they assume that the tracers follow nearly circular orbit. The advantage is that their method is independent from any density model, although it strongly depends  on values of Galactic constants (Sun’s distance from, and circular rotation speed around, the Galactic centre). Nevertheless, their results for rotation curve for various values of Galactic constants are consistent with our findings. We discuss these results more in detail in what follows. \\

\section{Density distribution from the Poisson equation}\label{ch4}
\begin{figure*}
        \begin{center}
                \subfloat[Fit of $a(z)$]
                {\includegraphics[width=0.5\textwidth]{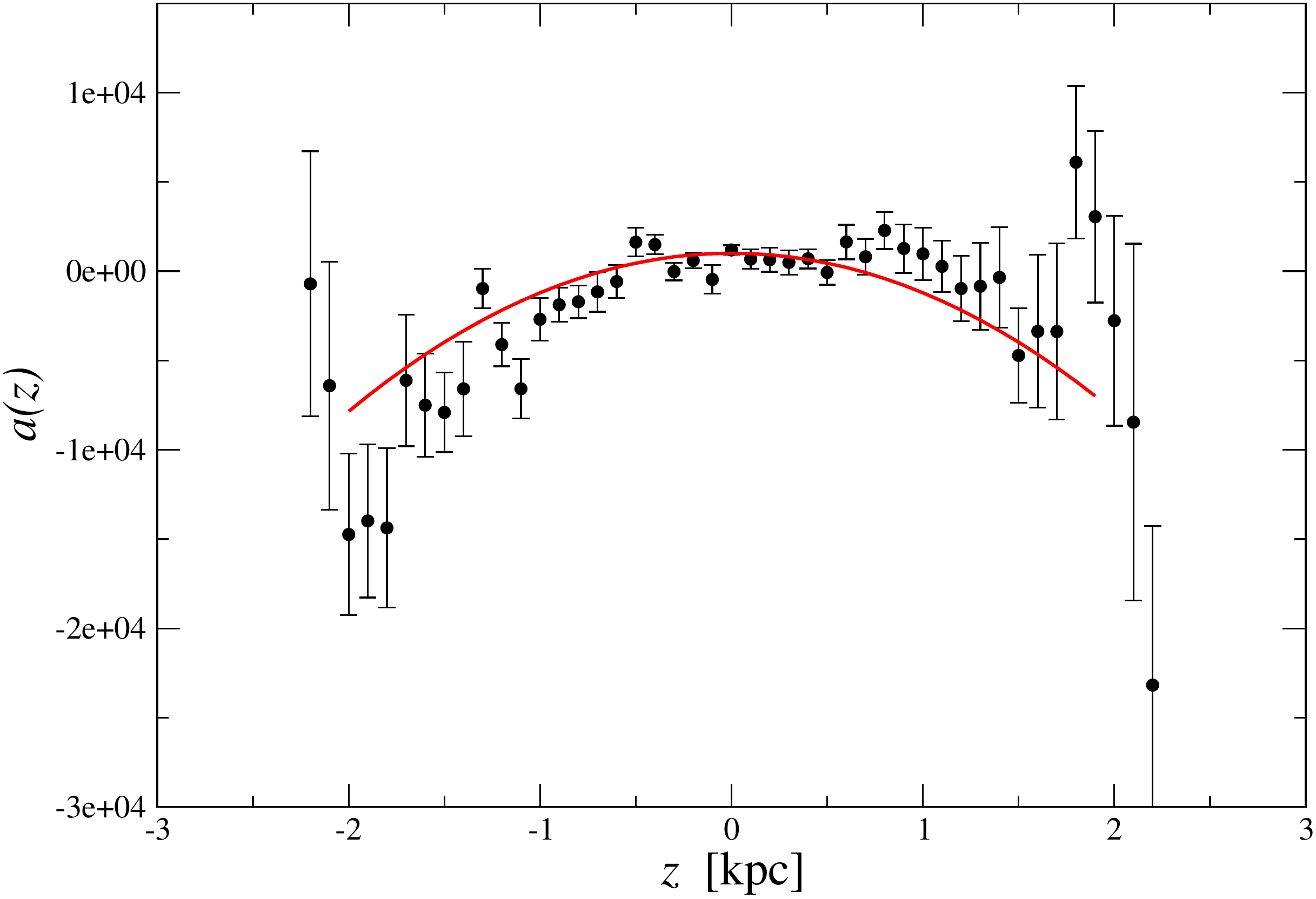}}
                \subfloat[Fit of $b(z)$]
                {\includegraphics[width=0.5\textwidth]{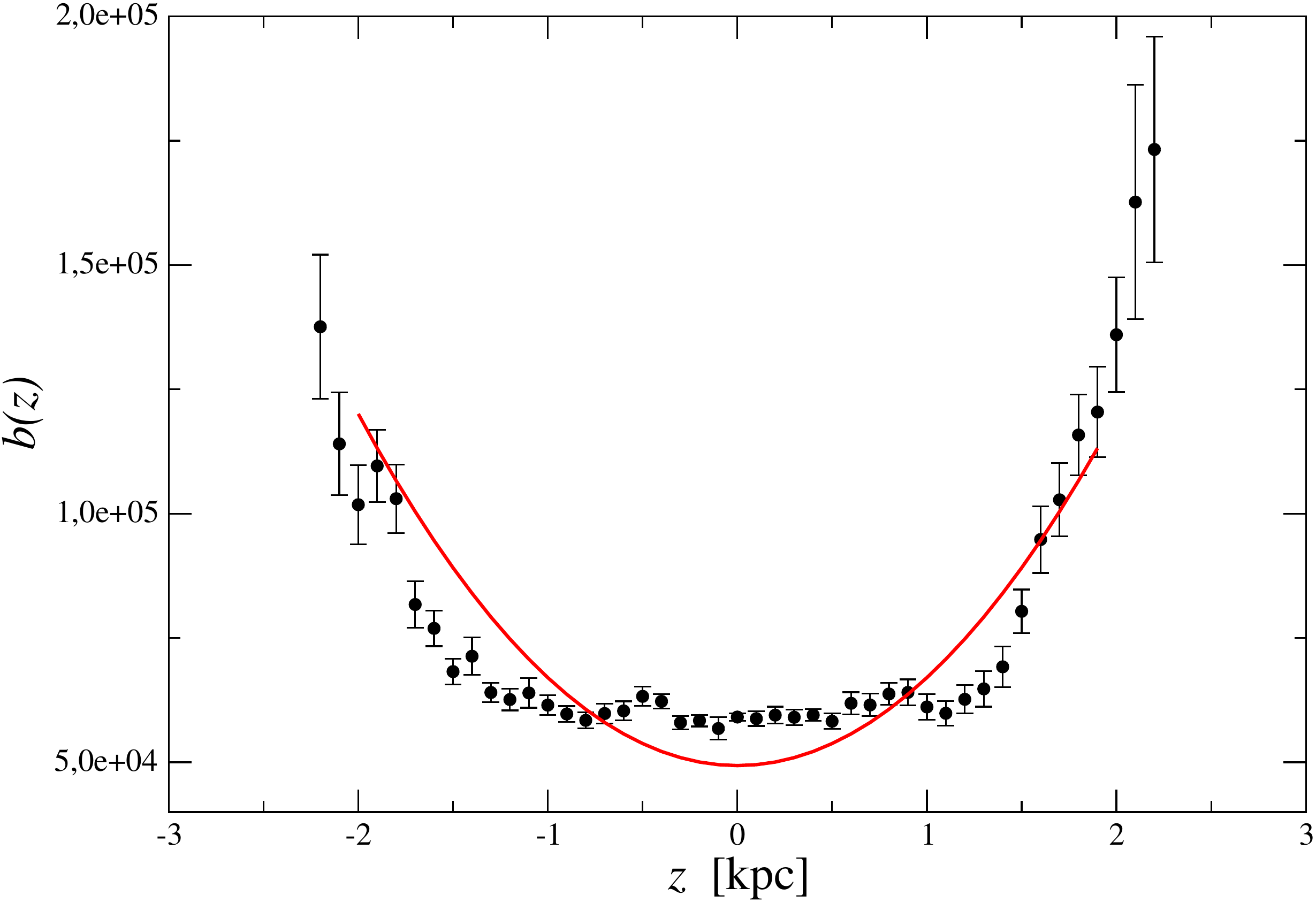}}
                \caption{ Fit of $a(z)$ and $b(z)$ (see Eq.\ref{fsl1}). 
                        The error bars represent the standard deviation.}
                \label{o4}
        \end{center}
\end{figure*}

Based on the results obtained for the rotation curve, we proceed to determine the density distribution in the Milky Way by considering different approaches. The first one is based on the Poisson equation in cylindrical coordinates and it assumes the dependence of the rotation speed with the azimuth to be negligible:

\begin{equation}\label{4}
\frac{1}{R}\frac{\partial}{\partial R}\left(R\frac{\partial\Phi}{\partial R}\right)+\frac{\partial^{2}\Phi}{\partial z^{2}}=4\pi G\rho(R,z)~.
\end{equation}
The first term on left side can be easily obtained by using Eq. (\ref{5a}). The second term, on the left side, can be obtained with the same relation and switching derivatives

\begin{eqnarray}\label{6}
\frac{\partial^2}{\partial z^2}\left(\frac{\partial\Phi}{\partial R}\right)&=& \nonumber \frac{1}{R}\frac{\partial^{2}v_{c}^{2}}{\partial z^{2}}~ \nonumber~, \\
\frac{\partial}{\partial R}\left(\frac{\partial^{2}\Phi}{\partial \nonumber z^{2}}\right)&=&\frac{1}{R}\frac{\partial^{2}v_{c}^{2}}{\partial z^{2}}~. 
\end{eqnarray}
By integrating the latter relation we find
\begin{eqnarray}
\frac{\partial^{2}\Phi}{\partial z^{2}}=-\int_{R}^{R_{max}} \frac{1}{R}\frac{\partial^{2}v_{c}^{2}}{\partial z^{2}} \mathrm{d}R + \Phi(R_{max},z=0)~. 
\end{eqnarray}
To determine derivatives of $v_c$ with respect to $R,z$, we assume that $v_{c}^2$ has a linear behaviour of the type
\begin{equation}\label{vc2}
v_{c}^2 = a(z)(R-14)+b(z) \,,
\end{equation}
where clearly $a(z)$ and $b(z)$  must be determined from the data. We find that $a(z)$ and $b(z)$ can be nicely fitted by parabolas and therefore, we can write
\begin{equation}
\label{fsl1} 
v_{c}^{2}(z) =\left[(\alpha+\beta z^{2})+(\gamma+\delta z^{2})(R-14)\right]\;,
\end{equation}
where the numerical values of $\alpha,\beta,\gamma,\delta$ are estimated from the data and the values  are given below. We use Eq. (\ref{5a}) and (\ref{vc2}) to express the first term of Eq. (\ref{4}) as
\begin{eqnarray}\label{7}
\frac{1}{R}\frac{\partial}{\partial R}\left(R\frac{\partial\Phi}{\partial R}\right)&=&{\frac{1}{R}\frac{\partial}{\partial R}\left(a(z)(R-14)+b(z)\right)} \nonumber \\
&=&\frac{a(z)}{R}
\end{eqnarray}
By making the derivative of the fit of the rotational velocity (Eq.\ref{fsl1}) with respect to $z,$ we express Eq. (\ref{6})  as

\begin{eqnarray}\label{8}
\frac{\partial^{2}\Phi}{\partial z^{2}}&=&2\beta\mathrm{ln}\left(\frac{R}{R_{max}}\right)+2\delta(R-R_{max}) \nonumber \\
&-&28\delta\mathrm{ln}\left(\frac{R}{R_{max}}\right)+\Phi(R_{max},z=0)~.
\end{eqnarray}
We find that the best fit values for $a(z)$ and $b(z)$ are (see Fig. \ref{o4}):

\begin{equation}\label{9}
\begin{split}
a(z)&=(-2200 \pm 400 )z^{2}+(1000 \pm 1000)  \\ 
b(z)&=(11400 \pm 1000 )z^{2}+(53000 \pm 1500) \;. 
\nonumber
\end{split}
\end{equation}
In the Galactic plane, the value of $a(z)$ is positive, which means that in the plane the 
velocity gradient is positive too: this must be compensated by density increase. That is clearly non-physical as we know that in our Galaxy, the density decreases exponentially in the outwards direction. Therefore, we conclude that we cannot use the Poisson equation to determine the density analytically. This problem may be related to large fluctuations present in the data, as well as by the fact that the system is not in equilibrium, so it does not satisfy the assumptions of the Jeans equation. We analyse the effect of the deviations from equilibrium in greater detail in Section \ref{ch6}.

\section{Density fit with the dark matter model}\label{ch5}
Another method to fit the rotation curve data can be done by making use of known density models. By assuming that the system is in equilibrium and made by different mass components, both the density and the rotational velocity can be expressed as
\begin{eqnarray}
\label{gal_exp}
\rho&=&\rho_{\mathrm{bulge}}+\rho_{\mathrm{disk}}+\rho_{\mathrm{halo}}~, \\
v_{c}^2 &=& v_{c,\mathrm{bulge}}^2+v_{c,\mathrm{disk}}^2+v_{c,\mathrm{halo}}^2~,
\end{eqnarray}
that is, we have decomposed the density and the circular velocity as the sum of three terms: the bulge, the disk, and the halo. Here, we examine each of these terms in more detail.

We do not fit the bulge, as we are interested mainly in outer parts of disk, where contribution of the bulge is negligible: we use
\begin{eqnarray}\label{17}
v_{c,bulge}^2=\frac{GM_{\mathrm{bulge}}}{R}~,
\end{eqnarray}
where  $M_{\mathrm{bulge}}=2\cdot10^{10}  M_{\odot}$ \citep{valenti}. 

For the disk, by assuming the balance between 
        the gravitational and centrifugal forces  
        at a generic point $(r,\phi,h)$
        (for a detailed derivation see Appendix \ref{odvodenie}) we derive: 
\begin{eqnarray}\label{16}      &&\mkern-20mu\int_{-H/2}^{H/2}\int_{R_{min}}^{R_{max}}\frac{2}{r}\left[\frac{(\hat{r}+r)(\hat{r}-r)+\Delta h^2}{[(\hat{r}-r)^2+\Delta h^2]\sqrt{(\hat{r}+r)^2+\Delta h^2}}E(k) \right. \nonumber \\
&&\mkern-20mu\left.-\frac{1}{\sqrt{(\hat{r}+r)^2+\Delta h^2}}K(k)\right] \nu(\hat{r},\hat{h})
\mathrm{d}\hat{r}\mathrm{d}\hat{h} \nonumber \\
&&\mkern-20mu+A\frac{v_{c,\mathrm{disk}}(r,h)^2}{r}=0~,
\end{eqnarray}
where $K(k),E(k)$ are  complete elliptic integrals of the first and second kind respectively, and
\begin{eqnarray}
k^2=\frac{4\hat{r}r}{(\hat{r}+r)^2+\Delta h^2}~,
\end{eqnarray}
where $\Delta h^2=(\hat{h}-h)^2$. For the sake of simplicity, we consider only a thin disk and we approximate $\Delta h \approx h$. For the density in Eq.\ref{16} we used the relation:
\begin{eqnarray}
\nu(\hat{r},\hat{z}) = \rho_0e^{-\hat{r}/h_{R}}e^{-\lvert \hat{h} \rvert/h_{z}}\;.
\end{eqnarray}
In Eq.\ref{16} the constant $A$ is the Galactic rotation number defined as
\begin{eqnarray}
A=\frac{R_{g,max}V_0^2}{G M_{d,max}}~,
\end{eqnarray}
where $M_{d,max}$ is mass of the disk, for which we use the value $M_{d,max}=6.5\cdot10^{10} M_\odot$ \citep{sofue_mass}. $R_{g,max}$ is the radius of the disk, which we fix at $25$ kpc, $V_0$ is the maximum velocity corresponding to the flat part of the rotation curve in the data-set: $257$ km/s in our case and $G$ is the gravitational constant: $4.302\cdot10^{-6}$ kpc $M_\odot^{-1}$ (km/s)$^2$.
We  calculate the fit in the Galactic plane, 
where $\Delta h \rightarrow 0$ and  Eq. (\ref{16}) becomes
\begin{eqnarray}
\label{14}
&&
\int_{R_{min}}^{R_{max}} \left[\frac{E(k)}{\hat{r}-r}-\frac{K(k)}{\hat{r}+r}\right]\rho_0e^{-\hat{r}/h_{R}}\hat{r} \mathrm{d}\hat{r} 
\\ \nonumber && 
+ A\frac{v_{c,\mathrm{disk}}(r)^2}{2\lvert h \rvert}=0~,
\end{eqnarray}
where 
\begin{eqnarray}
k^2=\frac{4r\hat{r}}{(\hat{r}+r)^2}~.
\end{eqnarray}

To fit the dark matter halo, we assume this is well approximated by the so-called Navarro, Frenk, and White density profile \citep{nfw}
\begin{eqnarray}
\label{nfw}
\rho_{\mathrm{halo}}&=&\frac{\rho_{0h}}{\frac{R}{R_s}\left(1+\frac{R}{R_s}\right)^2}~, \\
v_{c,halo}^2(R)&=&\frac{4\pi G \rho_{0h} R_s^3}{R}\left[\log \left(\frac{R_s+R}{R_s}\right)-\frac{R}{R_s+R}\right]~.
\end{eqnarray}
We use the least-squares method to find the best values of the free parameters. As this method requires a long computational time, we fix some well-known parameters and only fit those that are not so well determined. First, we fit only data in the Galactic plane, where we fix $h_R=2.5 $ kpc and $R_s=14.8 $ kpc, which are the values found by \cite{eilers}.
For the free parameters, we obtain the values $\rho_{0h}=2\cdot10^7 M_{\odot}/$kpc$^3$ and $\rho_0=3.83\cdot10^8 M_{\odot}/$kpc$^3$, with the value of the minimal $\chi^2=15.424$ for 107 points. 
We plot this result in Fig. \ref{o20} (a). We see that our rotation curves are well explained by a dominant dark matter halo, with a minimal contribution from the disc. From these values, we calculate the mass of the dark matter halo up to 25 kpc to be $M_h=3.52\cdot10^{11} M_{\odot}$, which is smaller than $7.25\cdot10^{11} M_{\odot}$ found by \cite{eilers}, but higher than $2.9\cdot 10^{11} M_\odot$ found by \cite{bajkova}. For the disk, we find $M_d=1.41\cdot10^{10} M_\odot$, which is lower than values found in the literature, for example, $6.5\cdot 10^{10} M_\odot$ as found by \cite{sofue_disk}, $0.95 \cdot 10^{11} M_\odot$ as found by \cite{kafle}, or $6.51\cdot 10^{10} M_\odot$ as found by \cite{bajkova}. \\

\begin{figure*}
        \vspace{-0.3cm}
        \centering
        \subfloat[$z=0$ kpc]{
                \includegraphics[width=0.3\textwidth]{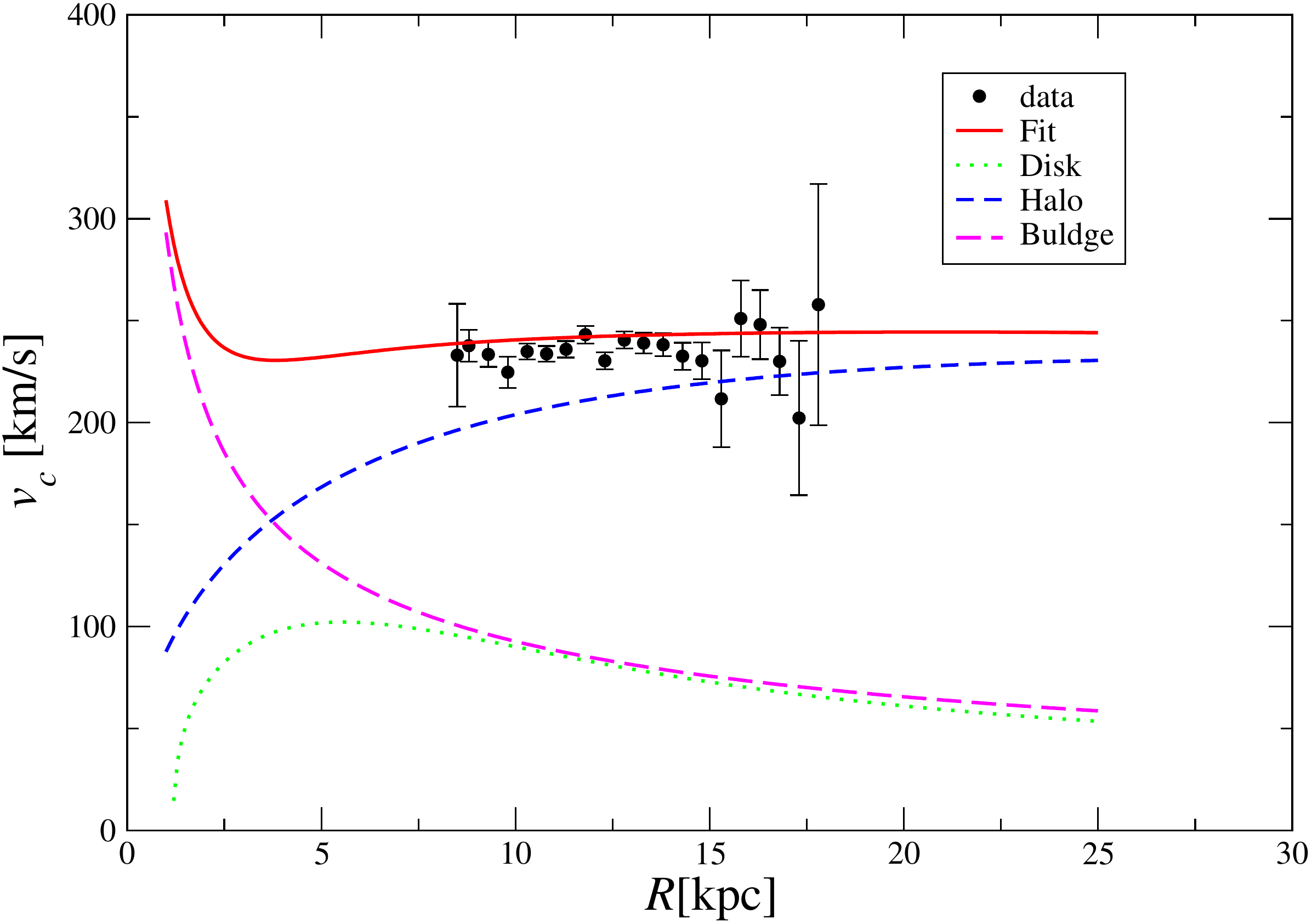}
        }
        \subfloat[$z=1.0$ kpc]{
                \includegraphics[width=0.3\textwidth]{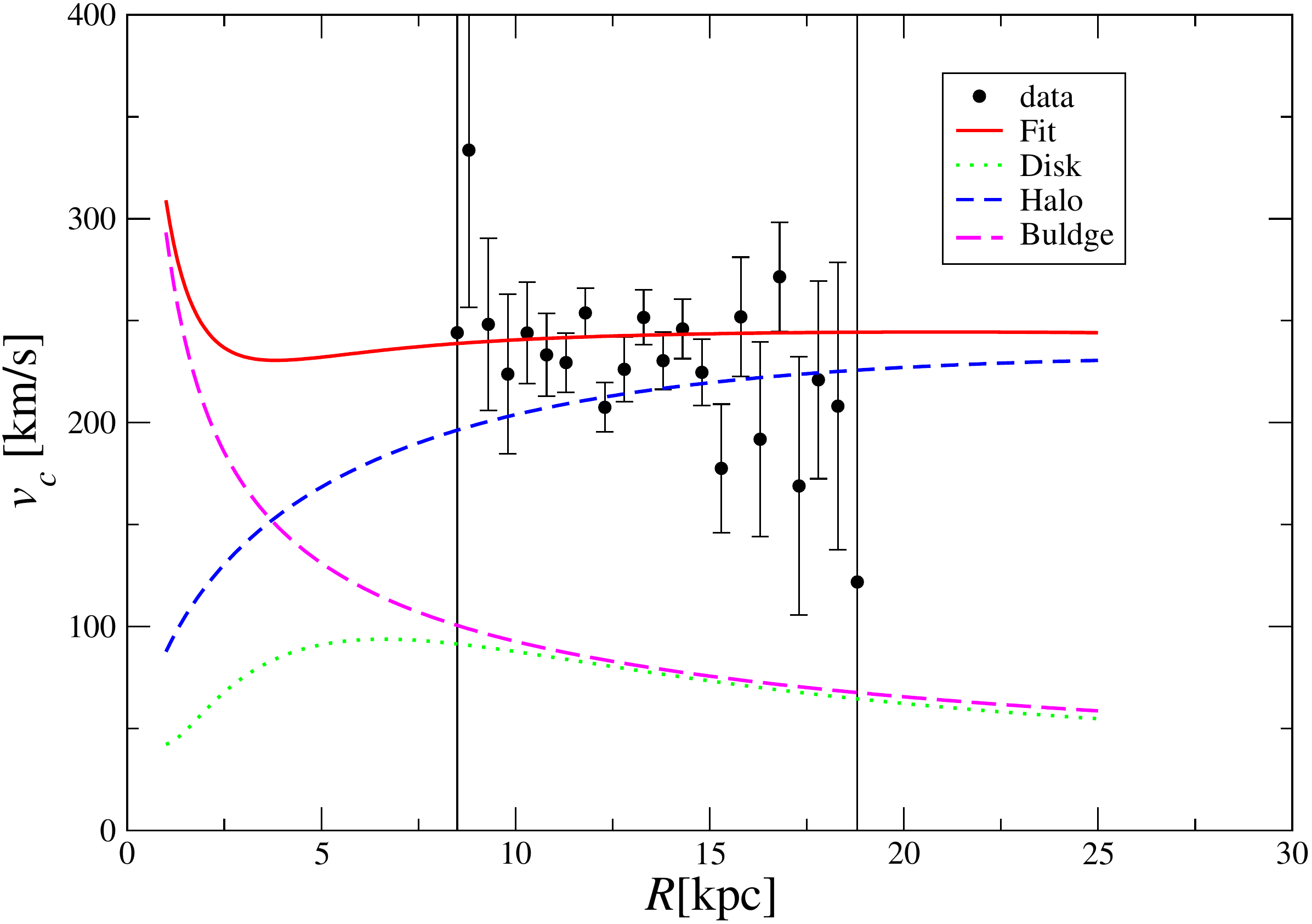}
        }
        \subfloat[$z=-1.0$ kpc]{
                \includegraphics[width=0.3\textwidth]{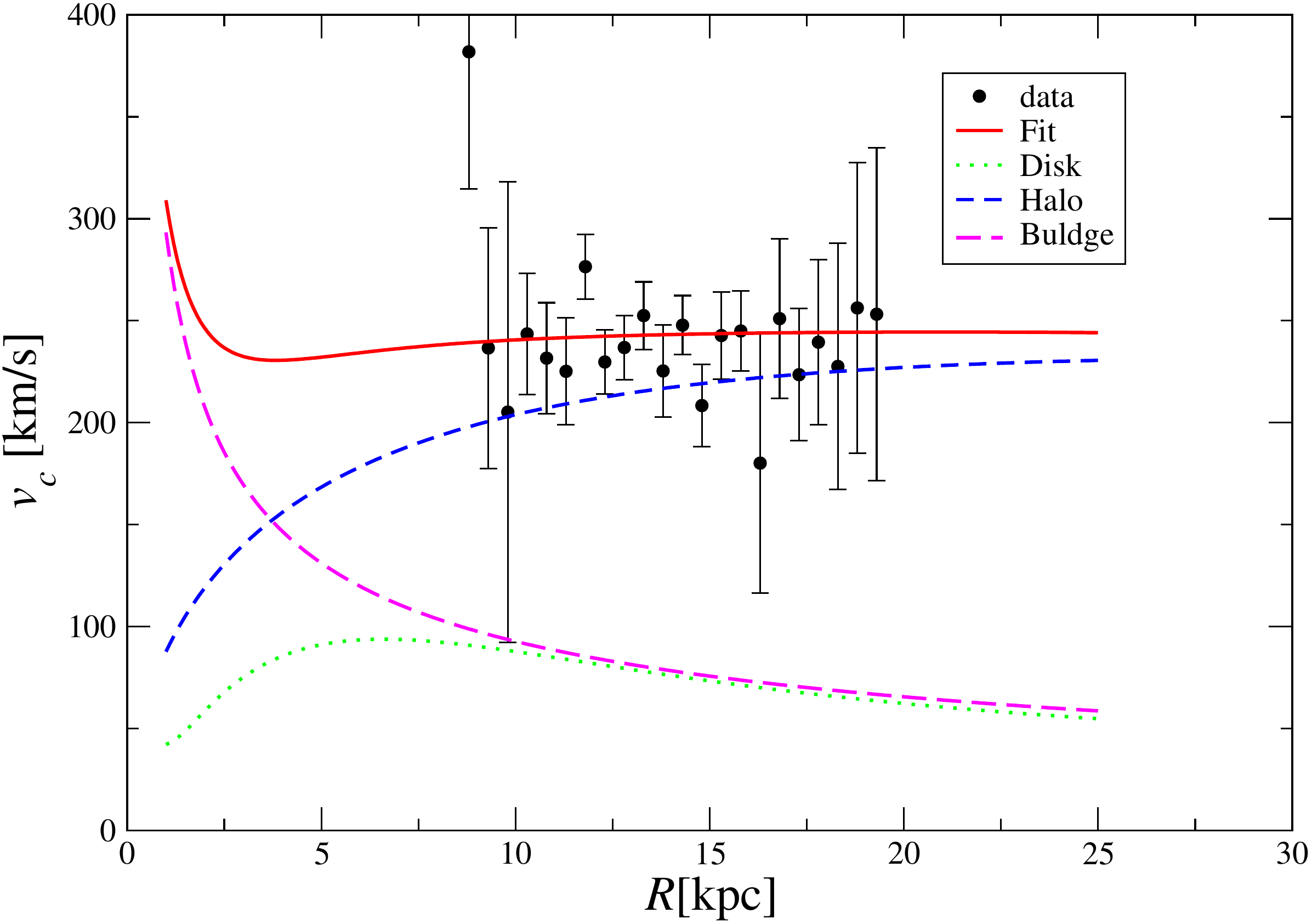}
        }
        \caption{Fit of the rotation curve for various values of $z$ using dark matter model.
                The data is binned with bins of size $\Delta R=0.5$ kpc and $\Delta z=0.1$ kpc}\label{o11}
\end{figure*}

For the off-plane data, we fit rotation curves for different values of $z$ at the same time, using relation (\ref{16}), which adds one more free parameter $h_z$ to the fit. Again, to save computational time, we restricted the number of free parameters and fixed $h_R=2.5$ kpc and $R_s=14.8$ kpc. We find $h_z=0.3 $ kpc, $\rho_0=4.1\cdot10^9 M_{\odot}/$kpc$^3$ and $\rho_{0h}=2.389\cdot10^7 M_{\odot}/$kpc$^3$, with the value of minimal $\chi^2=2510.37$ for 4653 points. In Fig. \ref{o11}, we plot the fit for various values of $z$. We see that in all cases, the dark matter halo is strongly dominant and the contribution from the disk is less important, which is as expected from rotational velocity that does not change with vertical distance. \\

This result is in agreement with result of \cite{eilers}, who also fitted their rotation curve with a similar model. They also find a dominant dark matter halo, with free parameter $\rho_{0h}=1.06\cdot10^7 M_{\odot}/$kpc$^3$. However, we disagree with result from \cite{jalocha}, who found that the gross mass distribution in our Galaxy is disk-like, without the need for an halo. 
\cite{jalocha} obtained their result based on modelling vertical gradient of azimuthal velocity, assuming the quasi-circular orbit approximation, and relating $v_c$ to $v_\phi$ directly from the balance condition of the radial component of gravitational and inertial force. We guess that the difference between our results comes from the fact that \cite{jalocha} did not take the Jeans equations into account when deriving rotational velocity. Indeed, in this latter work, assuming quasi-circular orbits  $v_\phi$ was directly related to  $v_c$ to obtain $v_\phi=\frac{r}{R} v_c$.

\section{Density fit with the MOND model}\label{ch_mond}
We tried to fit our results using the MOND theory, without invoking the presence of a heavy dark matter halo. To this purpose we have recalculated the expressions for the disk and the bulge, using relations from MOND \citep{milgrom_mond}:

\begin{eqnarray}\label{mond}
a_{M}=\frac{a_{N}}{\mu(\frac{a_{M}}{{a_0}})}~,
\end{eqnarray}
where
\begin{eqnarray}\label{15}
\mu\left(\frac{a_{{M}}}{a_0}\right)=
\sqrt{\frac{1}{1+\left(\frac{a_0}{a_{{M}}}\right)^2}}~,
\end{eqnarray}
with the value of $a_0=1.2\cdot 10^{-10}~\mathrm{m}\mathrm{s}^{-2}$ \citep{scarpa}. Solving Eq. (\ref{mond}) analytically yields

\begin{eqnarray}\label{18}
a_{M}=\sqrt{\frac{1}{2}a_{N}^2+\sqrt{\frac{1}{4}a_{N}^4+a_{N}^2a_0^2}}~.
\end{eqnarray}
Eq. (\ref{mond}) is indeed an approximation, which does not exactly stray from a spherical symmetric mass distribution. The exact solution may be analysed in the context for Bekenstein-Milgrom MOND theory derived from the modification of classical Newtonian dynamics \citep{brada}. However, the difference between the approximation of Eq. (\ref{mond}) and the exact solution is small, so we neglect it here. \\

\begin{figure*}
        \begin{center}
                \subfloat[Dark matter]
                {\includegraphics[width=0.5\textwidth]{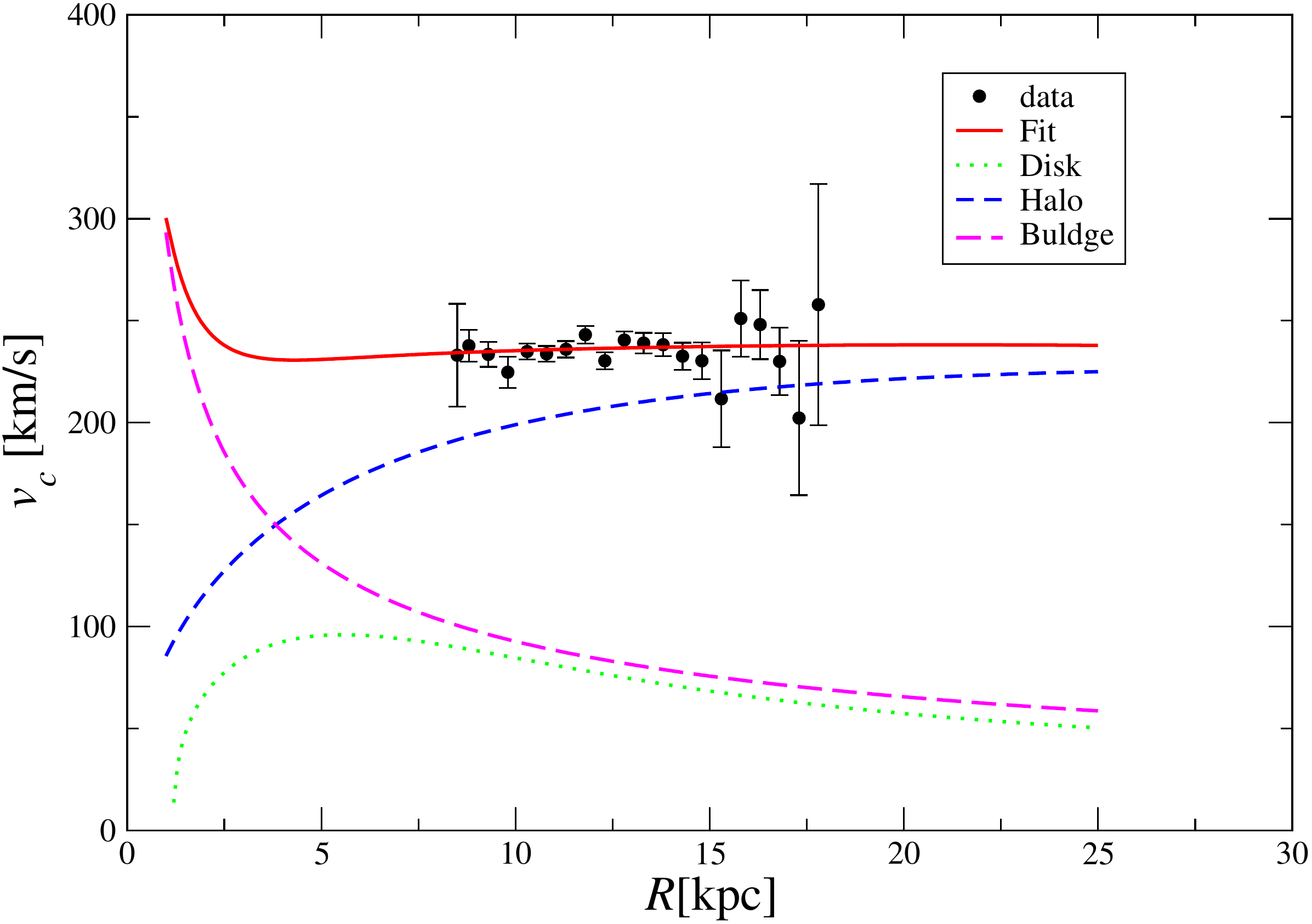}}
                \subfloat[MOND]
                {\includegraphics[width=0.5\textwidth]{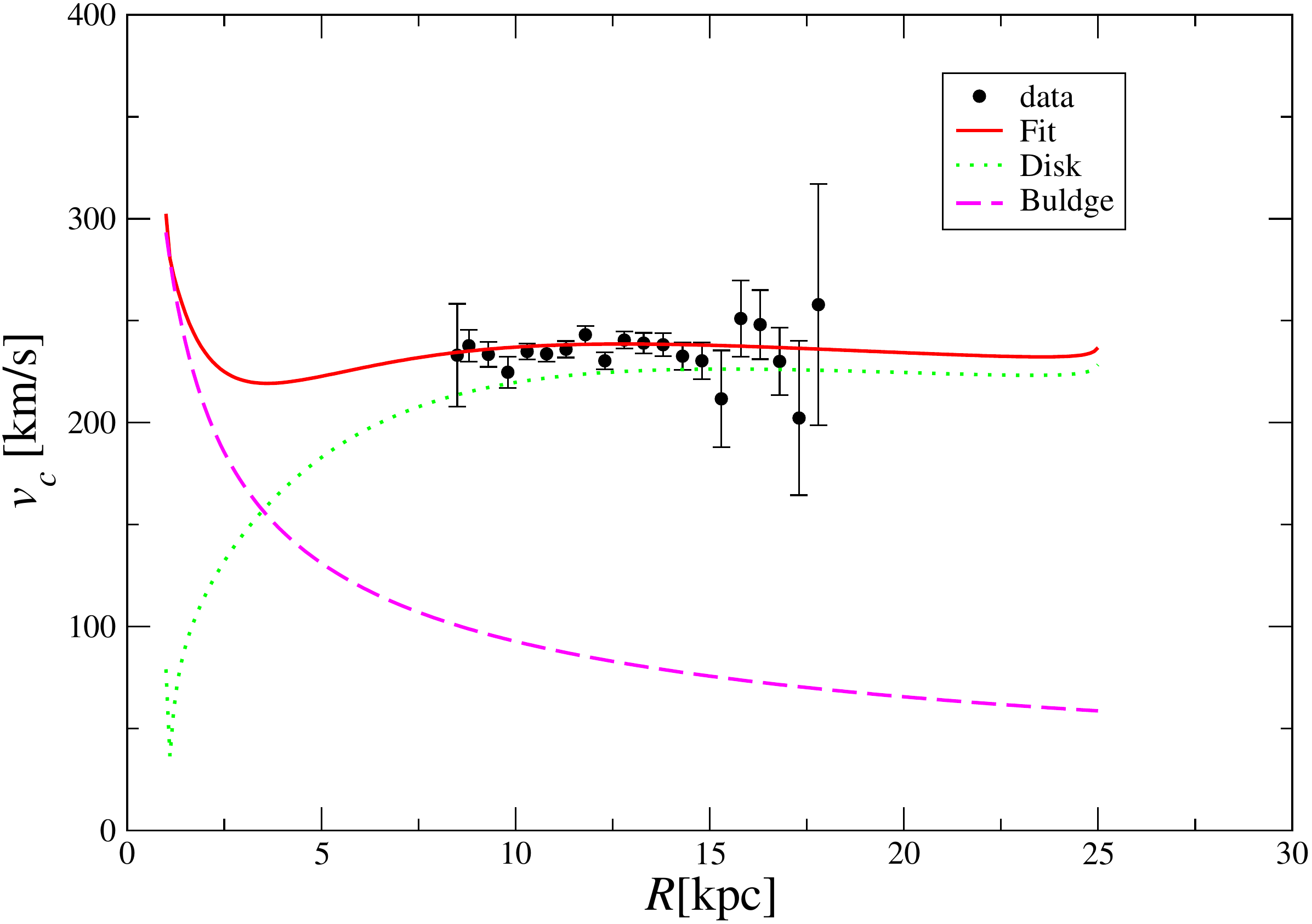}}
                \caption{Fit of the rotation curve for only $z=0$ using dark matter (left) and the  MOND theory (right). The data are
                        binned with bins of size $\Delta R=0.5$ kpc and $\Delta z=0.1$ kpc.}\label{o20}
        \end{center}
\end{figure*}

For the fit, we only used the disk and the bulge components. In Fig. \ref{o20} (b), we plot the result of the fit for the Galactic plane, nicely matching the observed value. For the free parameters, we found $\rho_0=7.49\cdot 10^8 M_{\odot}/$kpc$^3$ and $h_R=4.8$ kpc. The values of minimal $\chi^2$ is $\chi^2=15.776$ for 107 points, which is similar to the value for Newtonian fit. The mass of the disk up to 25 kpc found with these parameters is $M_d=2.77\cdot10^{10} M_\odot$ which is almost two times higher than that obtained with the dark matter model. \\

\begin{figure*}
        \vspace{-0.3cm}
        \centering
        \subfloat[$z=0$ kpc]{
                \includegraphics[width=0.3\textwidth]{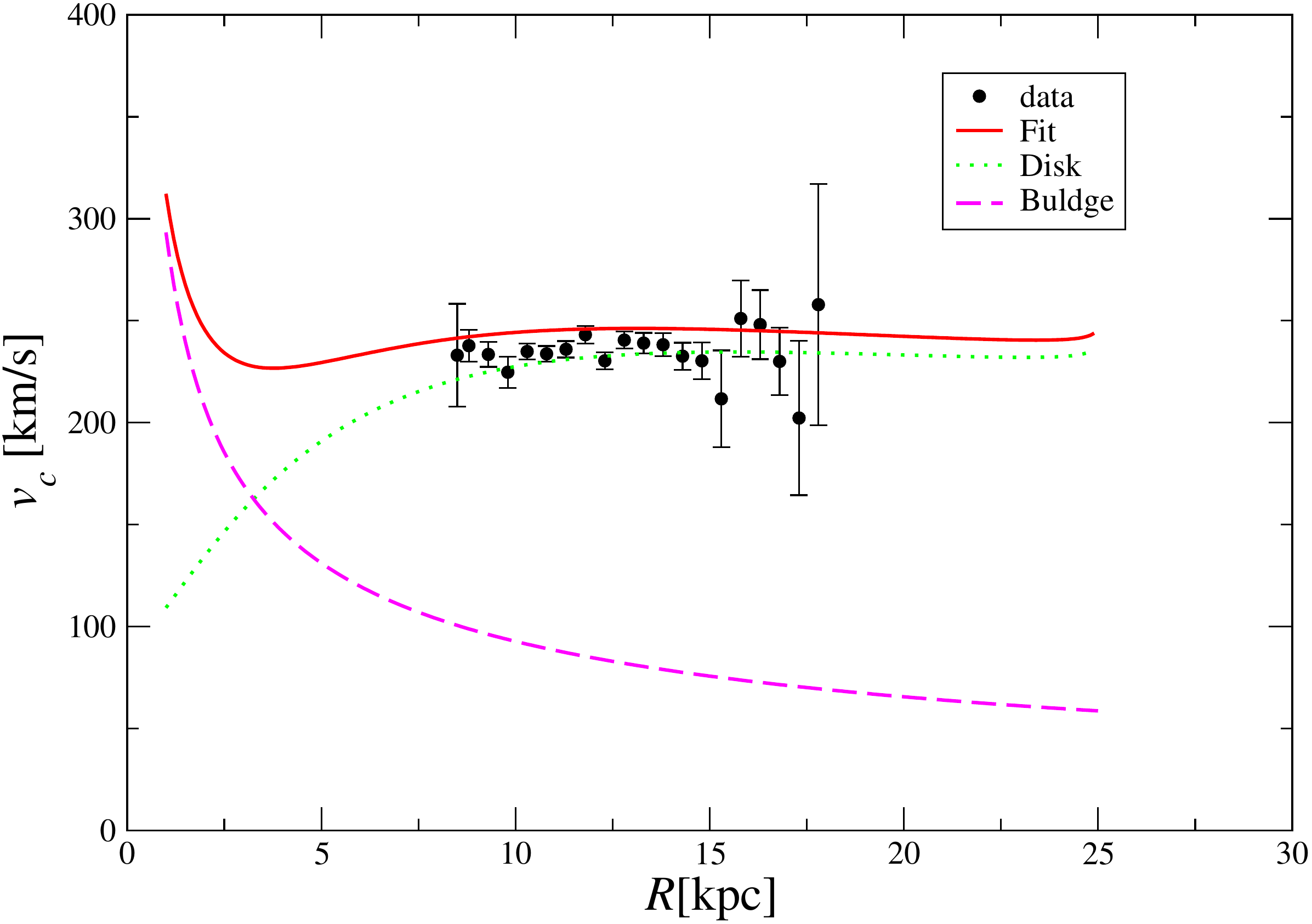}
        }
        \subfloat[$z=1.0$ kpc]{
                \includegraphics[width=0.3\textwidth]{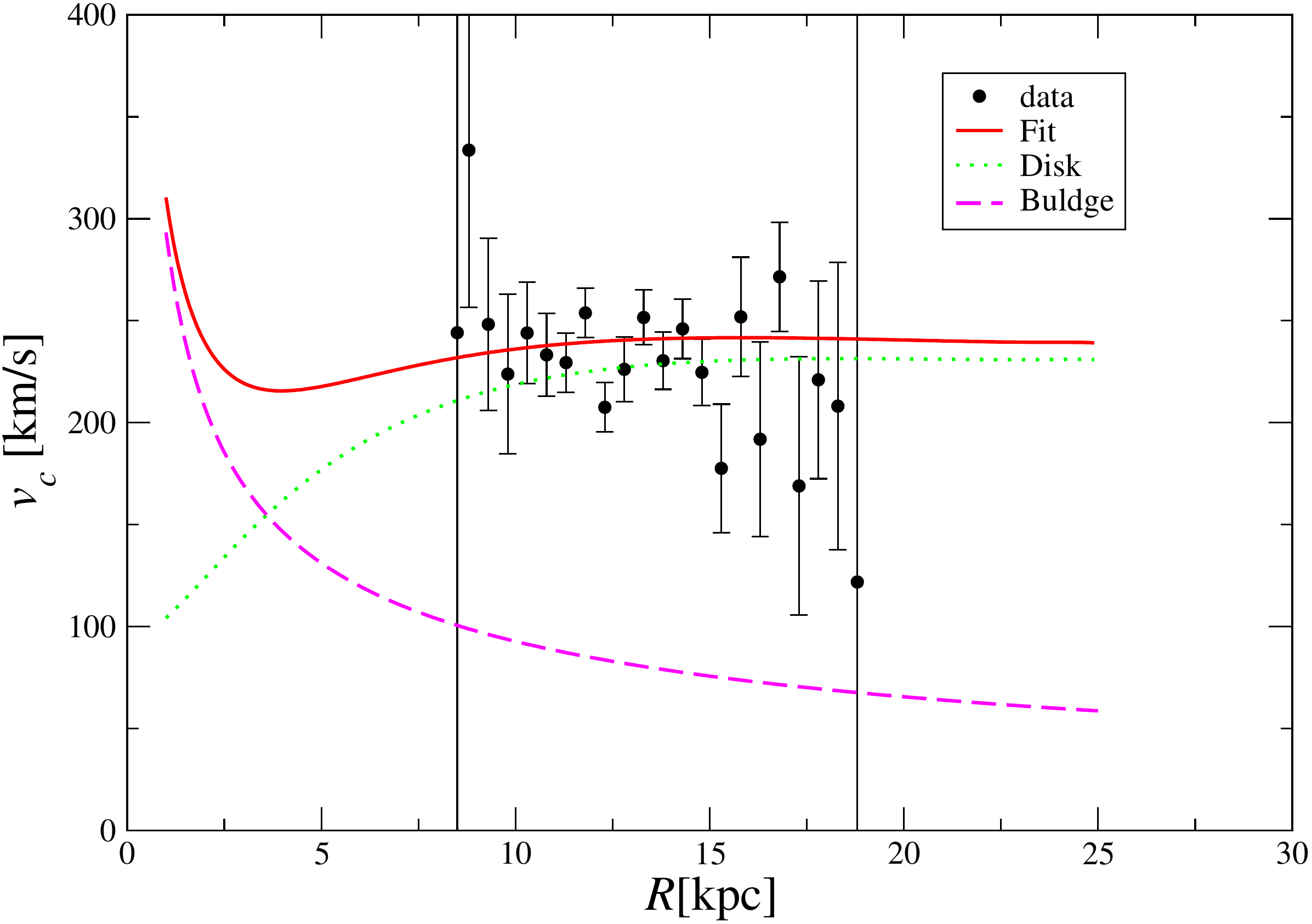}
        }
        \subfloat[$z=-1.0$ kpc]{
                \includegraphics[width=0.3\textwidth]{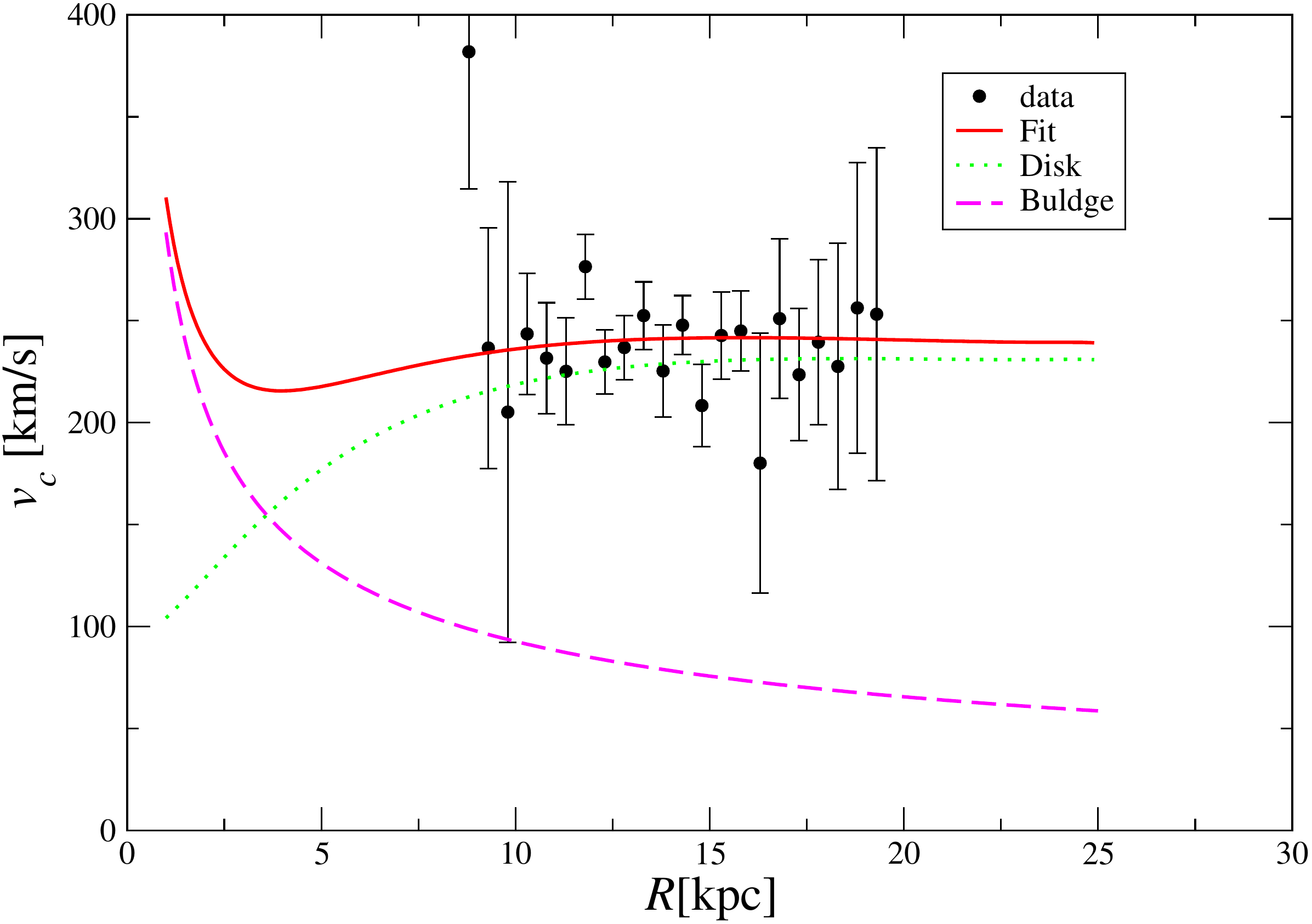}
        }
        \caption{Fit of the rotation curve for various values of $z$ using  
                the MOND theory. The data are binned with bins of size $\Delta R=0.5$ kpc and $\Delta z=0.1$ kpc.}\label{o19}
\end{figure*}

We tried to fit the off-plane rotation curve with the same approach. Again, we fit data for all $z$ with models for all $z$ at the same time. Thus, we find: $\rho_0=9.15\cdot10^9 M_{\odot}/$kpc$^3$ and $h_R=5.0$ kpc. We fixed the value of scale-height to $h_z=0.3$ kpc. The obtained value of minimal $\chi^2=2677.58$ for 4653 points, which is comparable with the Newtonian case. In Fig. \ref{o19}, we plot the results of the fit with MOND for different values of $z$. We see that off-plane, the fit is satisfying and there is no preference for the dark matter model over the MOND model. However, our result contradicts that of \cite{liasnti}, who also used Milky Way observables to compare the differences between dark matter and MOND theories. They performed a Bayesian likelihood analysis to compare the predictions of the model with the observed quantities. They find that the dark matter model is preferred, as MOND-like theories struggle to simultaneously explain both the rotational velocity and vertical motion of nearby stars in the Milky Way.

\section{Corrections to the Jeans equation}\label{ch6}
So far, we used the Jeans equation to determine the rotation curve of the Milky Way and its density profile. We recall that the basic assumptions of the Jeans equation are that the system is collisionless, axisymmetric, and in equilibrium. While the first condition represents a reasonable working hypothesis since collisional effects take place on much longer time scales than those that astrophysically relevant, the recent Gaia data have shown that the Milky Way is not in a stationary situation as there are large-scale gradients in all components of the velocity field and there are clear deviations from axisymmetry \citep{gaia_kin_maps,haifeng2,martin,haifeng1}. The dynamical origin of such features represents an open problem that has been explored by several authors \citep{antoja,binney_equil}. For instance, it has been concluded that the Galactic disk is still dynamically young and was last perturbed less than 1 Gyr ago, therefore modelling it as axisymmetric and in equilibrium is incorrect \citep{antoja}. The problem of reliability with regard to the Jeans equation was also studied by \cite{haines}, who analysed an N-body simulation of a stellar disk which had been perturbed by the recent passage of a dwarf galaxy and studied the surface density of the system based on the Jeans equation. They found that the Jeans equation gives reasonable results in over-dense regions, but fails in under-dense regions. Thus, the development of non-equilibrium methods for estimating the dynamical matter density locally and in the outer disk is necessary. \\

In order to test the effects of the deviations from a stationary configuration and axisymmetry on the Jeans equation, we consider N-body simulations of mock galactic systems that are not completely in an equilibrium configuration. The evolution of these systems was discussed in details in \cite{Benhaiem+Joyce+SylosLabini_2017,Benhaiem+SylosLabini+Joyce_2019,SylosLabini_RCD_DLP_2020}.We consider, hereafter, one of these systems, consisting of a thin, rotating, self-gravitating disk embedded in an ellipsoidal dark matter halo with an isotropic velocity dispersion. The inner regions of this system are very close to a stationary configuration, while the outer regions are progressively out-of-equilibrium. The signature of such a situation is represented by the behaviour of the radial velocity averaged in shells: at small distances from the centre this is close to zero, while at large enough distances, it becomes positive: the amplitude grows with the distance from the centre. \\

\begin{figure}
        \includegraphics[width=0.4\textwidth]{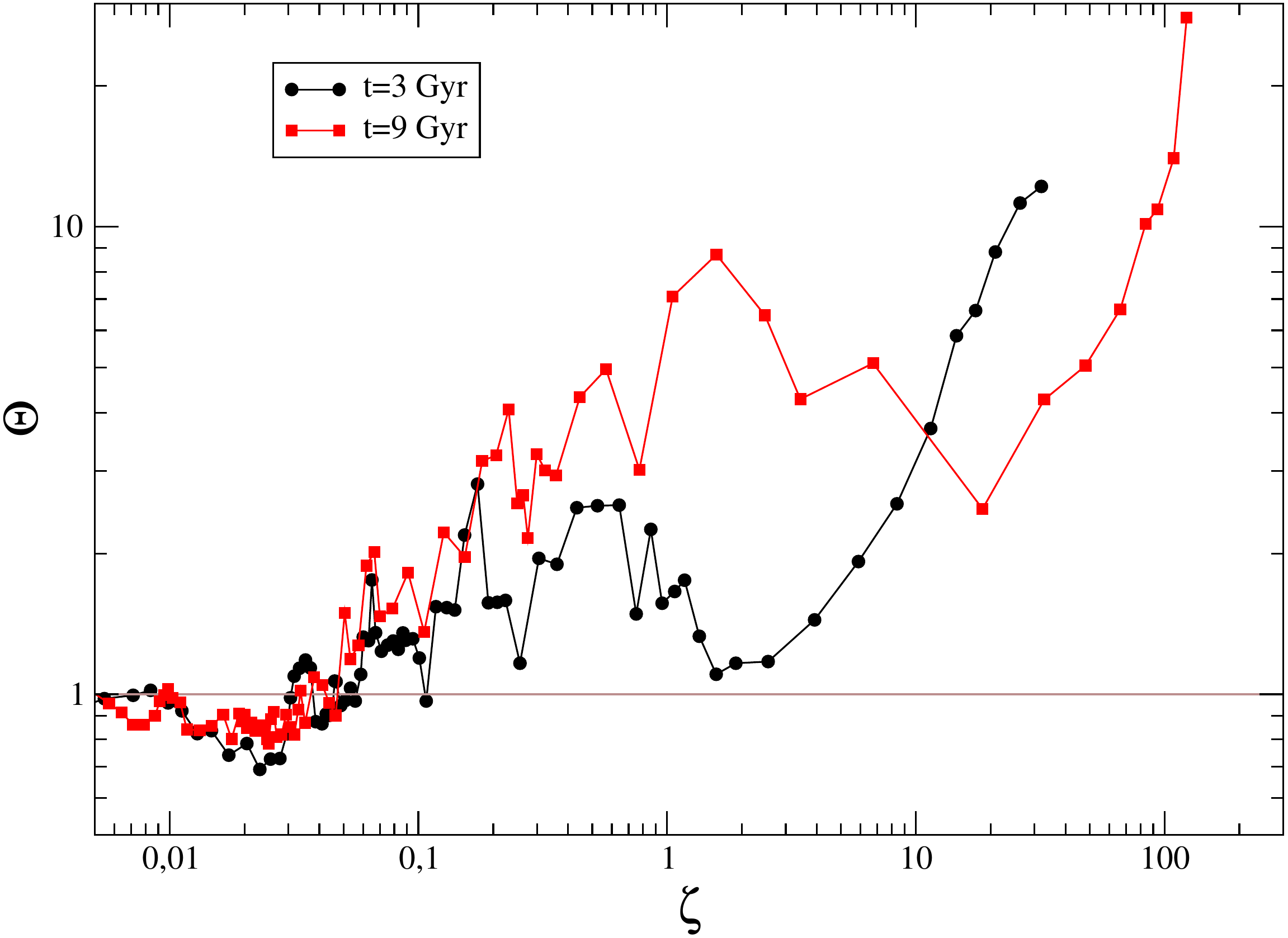}
        \caption{Ratio of the circular velocity from the Jeans equation and from the force
                (i.e. Eq.\ref{theta}) as a function of the ratio between the average radial velocity and the 
                azimuthal velocity (see Eq.\ref{zeta}). Black and red circles correspond 
                to the system evolved up to 3/9 Gyr, respectively.}\label{sila}
\end{figure}

The circular velocity from the Jeans equation is

\begin{eqnarray}\label{jeans_sim}
v_{c,J}^{2}=\overline{v_{\Phi}}^{2}-\overline{v_{R}}^{2}\left(1+\frac{\partial \mathrm{ln}\nu}{\partial \mathrm{ln}R}+\frac{\partial\mathrm{ln}\overline{v_{R}}^2}{\partial \mathrm{ln}R}\right)~,
\end{eqnarray}  
where we neglect the cross-term $\overline{v_Rv_\phi}$, as it's contribution to the final result is negligible $(\sim1\%)$ \citep{eilers}. By definition, the circular velocity can be computed from the gravitational force:

        \begin{eqnarray}
        v_{c,F}^{2}=R F_R= \left| \overline{\vec{F}} \cdot \vec{R} \right|
        ,\end{eqnarray}  
        where  $\overline{\vec{F}}$ is the gravitational force
        acting of the particles 
        contained in the two-dimensional corona at a distance, $R,$ and thickness, 
        $\Delta R$ (where $R$ is the cylindrical coordinate).  
        Thus, we compute the gravitational force  acting of the $i^{th}$
        particle as 
        \begin{eqnarray}
        \vec{F}_i =  G \sum_{j=1}^N m_j m_i
        \frac{(\vec{r}_i - \vec{r}_j)}{|\vec{r}_i - \vec{r}_j|^3} \;,
        \end{eqnarray}  
        where $m_i$ is the mass of the $i^{th}$ particle 
        and we compute its average in a corona. 
If axisymmetry and stationary equilibrium are established, then $v_{c,F} =v_{c,J}$: the difference between these two quantities thus depends on the deviations from the assumptions underlying the Jeans equation. In the Fig. \ref{sila}, we plot the ratio:

\begin{equation}
\label{theta}
\Theta = \frac{v_{c,J}}{v_{c,F}}
\end{equation}
as a function of 
\begin{equation}
\label{zeta} 
\zeta = \frac{|v_R|}{|v_\phi|}~.
\end{equation}
When the radial velocity is small, that is, $\zeta \ll 1,$ then $ \Theta \approx 1$, whereas when the radial velocity becomes larger than $10 \%$ of the azimuthal one, then $\Theta$ becomes larger than one. In the Milky Way, $\zeta \approx 0.1$ at $R \approx 20$ kpc, as found by LS19. We note that in Fig. \ref{sila}, we have reported the behaviour for two different times, that is, 3 and 9 Gyr; indeed, as the external regions are out-of-equilibrium they continue to evolve over time, while the inner regions are quasi stationary.

We conclude that the Jeans equation is reliable 
        when the radial velocity is smaller than 10\% of the azimuthal one, otherwise corrections to the Jeans equation
        become necessary. In particular, we find that 
        the estimation of the circular velocity 
        though the Jeans equation gives an overestimation
        with respect to the estimation
        of the circular velocity through the force. 
        This implies that by using $v_{c,J}$ to compute the mass 
        through the relation,
        \begin{equation}
        M_J(r) = \frac{v_{c,J}^2\times r}{G}
        ,\end{equation}
        the real mass is overestimated by a factor that is proportional to $\Theta^2$.

\section{Conclusions}
In this paper, we study the rotation curve of the Milky Way from the extended kinematic maps of Gaia-DR2. We calculated the rotation curve in plane and in off-plane regions, using the Jeans equation. Our results show that the rotation curve in the outer disk has very little dependence on $R$ and $z$. \\

We fitted the rotation curve using models with dark matter halo or MOND, using the least-squares method. We find that a model based on dark matter fits the data very well, and the results are in good agreement with other works. For the dark matter model, we obtain the minimal $\chi^2= 15.424$ for 107 points in the plane and $\chi^2=2510.37$ for 4653 points off plane. The MOND model in the plane gives $\chi^2= 15.776$ for 107 points, which is comparable with the dark matter model. Off-plane the results are similar as well, with $\chi^2=2677.58$ for 4653 points, which fits the data similarly to the dark matter model. \\

We also considered the corrections to  the Jeans equation in non-equilibrium and non-axisymmetric systems. Indeed, the Jeans equation assumes that the system is axisymmetric and in equilibrium, which is not the case of the Milky Way. For this reason, we consider N-body simulations of galaxies and calculated the rotational velocity by using the Jeans equation $v_{c,J}$ and by computing the gradient of the gravitational potential $v_{c,F}$. We find that the two ways of calculating the rotational velocity are in good agreement as long as the ratio, $\zeta,$ between the modulus of the radial velocity and of the azimuthal velocity is smaller than $\sim 10\%$. When  $\zeta$  becomes larger than this value, then $v_{c,J} > v_{c,F}$ and, thus, we overestimate the Galactic mass if we use the rotational velocity computed through the Jeans equation. For the case of the Milky Way, it was found in LS19 that $\zeta \approx 0.1$ at $R \approx 20$ kpc: this implies that at a larger galactocentric distance, using the Jeans equation leads to an overestimation the mass of the Milky Way. The Gaia DR3 will clarify whether in the range of distances $20 < R < 30$ kpc, such corrections may become large enough to change our view of the Galaxy as a quasi-equilibrium system, thus altering its estimated mass.

\begin{acknowledgements}
We thank the anonymous referee for helpful comments, which improved this paper and Agnes Monod-Gayraud (language editor of A\&A) for proof-reading of the text. ZC and MLC were supported by the grant PGC-2018-102249-B-100 of the Spanish Ministry of Economy and Competitiveness (MINECO). HFW is supported by the LAMOST Fellow project, National Key Basic R\&D Program of China via 2019YFA0405500 and funded by China Postdoctoral Science Foundation via grant 2019M653504, Yunnan province postdoctoral Directed culture Foundation and the Cultivation Project for LAMOST Scientific Payoff and Research Achievement of CAMS-CAS. RN was supported by the Scientific Grant Agency VEGA No. 1/0911/17. This work has made use of data from the European Space Agency (ESA) mission {\it Gaia} (\url{https://www.cosmos.esa.int/gaia}), processed by the {\it Gaia} Data Processing and Analysis Consortium (DPAC, \url{https://www.cosmos.esa.int/web/gaia/dpac/consortium}). Funding for the DPAC has been provided by national institutions, in particular the institutions participating in the {\it Gaia} Multilateral Agreement.
\end{acknowledgements}

\clearpage
\bibliographystyle{aa} 
\bibliography{Refer}

\appendix

\section{Derivation of the integral of the thin disk }\label{odvodenie}
To derive the rotation curve of a disk in 3D, we derive the equation for balance between the gravitational and the centrifugal force. We consider two points with the coordinates $P(r,\theta,z)$ and $Q(\hat{r},\hat{\theta},\hat{z})$. The distance between these two points can be expressed as $(\hat{r}^2-r^2-2r\hat{r}cos\hat{\theta}+\Delta h^2)^{1/2}$ and the vector projection as $(\hat{r}cos\hat{\theta}-r)$, where $\Delta h$ is the difference in heights $\Delta h=(\hat{h}-h)$. The Newtonian gravitational force on the point $P$ from a body consisting of points $Q$ distributed with a mass density $\hat{\rho}(\hat{r},\hat{h})$ can be expressed as an integral over these points:

\begin{eqnarray}
F_X\mkern-10mu &=&\mkern-10mu \dfrac{G M_g}{R_g^2} \int_{-H/2}^{H/2} \int_{0}^{2\pi} \int_{0}^{1}    \dfrac{\hat{r}\cos \hat{\theta} - r}{ ( \hat{r}^2 + r^2 - 2\hat{r}r \cos \hat{\theta} + \Delta h^2)^{3/2}} \nonumber \\ 
&\cdot&\mkern-15mu\hat{\rho}(\hat{r},\hat{h}) \hat{r} \spaceI d\hat{r} d\hat{\theta} d \hat{h}~.
\end{eqnarray}
The centrifugal force can be written simply as

\begin{equation}\label{CentrifugalForce}
F_c = \dfrac{V^2}{ R }= \dfrac{V_0^2}{R_g} \dfrac{v(r,h)^2}{r}~.
\end{equation}\ 
Here, we made all the variables dimensionless by measuring the distances in units of the
outermost galactic radius $R_g$, mass density $\hat{\rho}$ in units of $M_g/R_g^3$, where $M_g$ is the total galactic mass and velocities in units of the characteristic velocity $V_0$. So the balance between the gravitational and centrifugal force yields

\begin{eqnarray}\label{a1}
&&\mkern-40mu \int_{-H/2}^{H/2} \int_{0}^{2\pi} \int_{0}^{1}  \frac{\hat{r}cos\hat{\theta}-r}{(\hat{r}^2-r^2-2r\hat{r}cos\hat{\theta}+\Delta h^2)^{3/2}}\hat{\rho}(\hat{r},\hat{h})\hat{r}d\hat{r}d\hat{\theta}d\hat{h} \nonumber \\
&+&A\frac{v(r,h)^2}{r}=0~,
\end{eqnarray}
where $A$ is the galactic rotation number

\begin{eqnarray}
A=\frac{R_{g}V_0^2}{G M_{g}}~.
\end{eqnarray}
We get rid of $\hat{\theta}$ dependency by simplifying the integral

\begin{eqnarray}\label{a2}
I(r,\hat{r},\Delta h)=\int_{0}^{2\pi} \frac{\hat{r}cos\hat{\theta}-r}{(\hat{r}^2-r^2-2r\hat{r}cos\hat{\theta}+\Delta h^2)^{3/2}}d\hat{\theta}
\end{eqnarray}
using complete elliptic integrals of first and second kind. \citet[pages 179 \& 182]{integral} give the solution to these integrals

\begin{eqnarray}
I_1 =  \int \dfrac{ d x }{ ( a - b\cos x )^{1/2} }  & = & \dfrac{2}{\sqrt{ a+b }} F(\delta,k)~; \\
I_3 = \int \dfrac{ d x }{ ( a - b\cos x )^{3/2} } & = & \dfrac{2}{ (a-b) \sqrt{a+b} } E(\delta,k)~ ,
\end{eqnarray}
where

\begin{eqnarray}
&&\mkern-28mu x\in  [0,\pi]; \spaceII \sin \delta = \sqrt{ \dfrac{(a+b)(1-\cos \Phi)}{ 2( a - b \cos \Phi ) } }; \spaceII \\ &&\mkern-28mu k=\sqrt{\dfrac{2b}{a+b}}; \spaceII a>b>0; \spaceII \Phi \in [0,\pi]~.
\end{eqnarray}
$F(\delta,k)$ and $E(\delta,k)$ are the incomplete elliptic integrals of the first and second kind

\begin{eqnarray}\label{EllipticIntegrals}
F( \delta, k) &=& \int_0^{\delta} \dfrac{d \phi}{\sqrt{ 1 - k^2 \sin^2 \phi }}; \spaceII \nonumber\\
E( \delta, k) &=& \int_0^{\delta} \sqrt{ 1 - k^2 \sin^2 \phi } \spaceI d\phi~.
\end{eqnarray}
For the angle $\delta=\pi/2,$ we obtain complete elliptic integrals that we can rewrite by substituting $t=\sin \phi$ as

\begin{eqnarray}\label{EllipticIntegralsComplete}
K(k) &\equiv& F \left( \frac{\pi}{2} , k \right) = \int_0^1 \dfrac{dt}{\sqrt{ (1-t^2)(1 - k^2 t^2) }}; \nonumber \\
E(k) &\equiv& E \left( \frac{\pi}{2},k \right) = \int_0^1 \sqrt{ \dfrac{1 - k^2 t^2}{1-t^2} } \spaceI dt \spaceI .
\end{eqnarray}
When we plug our values:

\begin{eqnarray}
&\mkern-10mu a=r^2+\hat{r}^2+\Delta h^2; \spaceII b=2r\hat{r}
\end{eqnarray}
to the Eq. \ref{a1}, we get Eq.(\ref{16}):

\begin{eqnarray}        &&\mkern-20mu\int_{-H/2}^{H/2}\int_{R_{min}}^{R_{max}}\frac{2}{r}\left[\frac{(\hat{r}+r)(\hat{r}-r)+\Delta h^2}{[(\hat{r}-r)^2+\Delta h^2]\sqrt{(\hat{r}+r)^2+\Delta h^2}}E(k) \right. \nonumber \\
&&\mkern-20mu\left.-\frac{1}{\sqrt{(\hat{r}+r)^2+\Delta h^2}}K(k)\right]\rho_0e^{-\hat{r}/h_{R}}e^{-\lvert h \rvert/h_{z}}\hat{r}~ \mathrm{d}\hat{r}\mathrm{d}h \nonumber \\
&&\mkern-20mu+A\frac{v_{c,\mathrm{disk}}(r,h)^2}{r}=0~. \nonumber
\end{eqnarray}

\end{document}